\documentstyle[12pt,epsfig]{article}
\def\be{\begin{equation}}
\def\ee{\end{equation}}
\def\bc{\begin{center}}
\def\ec{\end{center}}
\def\bea{\begin{eqnarray}}
\def\eea{\end{eqnarray}}

\def\ev{{\rm \; eV}}
\def\gev{{\rm \; GeV}}
\def\h1b{{\overline{H}_1}}
\def\h2b{{\overline{H}_2}}

\def\kh{{\hat{K}}}
\def\kt{{\tilde{K}}}
\def\htil{{\tilde{H}}}
\def\wh{{\hat{w}}}
\def\wt{{\tilde{w}}}
\def\fh{{\hat{f}}}
\def\ft{{\tilde{f}}}

\def\l1{{\Lambda_1}}

\def\lam{{\Lambda}}

\def\lsu{{\Lambda}_{S}}
\def\mpl{M_{\rm P}}
\def\ov{\overline}
\def\simlt{\stackrel{<}{{}_\sim}}
\def\simgt{\stackrel{>}{{}_\sim}}
\def\str{{\rm Str \;}}

\newcommand{\nsect}{\setcounter{equation}{0}
\def\theequation{\thesection.\arabic{equation}}\section}

\newcommand{\appendixA}{\setcounter{equation}{0}
\def\theequation{\rm{A}.\arabic{equation}}\section*}

\catcode`@=11
\def\marginnote#1{}
\newcount\hour
\newcount\minute
\newtoks\amorpm
\hour=\time\divide\hour by60
\minute=\time{\multiply\hour by60 \global\advance\minute by-\hour}
\edef\standardtime{{\ifnum\hour<12 \global\amorpm={am}%
        \else\global\amorpm={pm}\advance\hour by-12 \fi
        \ifnum\hour=0 \hour=12 \fi
        \number\hour:\ifnum\minute<10 0\fi\number\minute\the\amorpm}}
\edef\militarytime{\number\hour:\ifnum\minute<10 0\fi\number\minute}
\def\draftlabel#1{{\@bsphack\if@filesw {\let\thepage\relax
   \xdef\@gtempa{\write\@auxout{\string
      \newlabel{#1}{{\@currentlabel}{\thepage}}}}}\@gtempa
   \if@nobreak \ifvmode\nobreak\fi\fi\fi\@esphack}
        \gdef\@eqnlabel{#1}}
\def\@eqnlabel{}
\def\@vacuum{}
\def\draftmarginnote#1{\marginpar{\raggedright\scriptsize\tt#1}}
\def\draft{\oddsidemargin 0.0truein
        \def\@oddfoot{\sl preliminary draft \hfil
        \rm\thepage\hfil\sl\today\quad\militarytime}
        \let\@evenfoot\@oddfoot \overfullrule 3pt
        \let\label=\draftlabel
        \let\marginnote=\draftmarginnote
   \def\@eqnnum{(\theequation)\rlap{\kern\marginparsep\tt\@eqnlabel}%
\global\let\@eqnlabel\@vacuum}  }
\catcode`@=12
\begin{document}
\begin{titlepage}
\vspace*{-1cm}
hep-ph/9703286
\hfill{CERN-TH/96-369}
\\
\phantom{bla}
\hfill{DFPD~96/TH/68}
\\
\vskip 1.0cm
\begin{center}
{\Large\bf Aspects of spontaneously broken $N=1$ global
supersymmetry in the presence of gauge interactions}
\footnote{Work supported in part by the European Union 
TMR project ERBFMRX-CT96-0045}
\end{center}
\vskip 0.8  cm
\begin{center}
{\large Andrea
Brignole}\footnote{e-mail address: brignole@vxcern.cern.ch}
\\
\vskip .1cm
Theory Division, CERN, CH-1211 Geneva 23, Switzerland
\\
\vskip .2cm
{\large Ferruccio
Feruglio}\footnote{e-mail address: feruglio@padova.infn.it}
\\
\vskip .1cm
Dipartimento di Fisica, Universit\`a di Padova, I-35131 Padua, Italy
\\
\vskip .2cm
and
\\
\vskip .2cm
{\large Fabio
Zwirner}\footnote{e-mail address: zwirner@padova.infn.it}
\\
\vskip .1cm
INFN, Sezione di Padova, I-35131 Padua, Italy
\end{center}
\vskip 0.5cm
\begin{abstract}
\noindent
We discuss models where $N=1$ global supersymmetry is
spontaneously broken, at the classical level, in the
presence of non-anomalous gauge interactions. We take
such models as effective theories, valid up to some 
suitable scale and arising from supergravity models 
with a light gravitino, and therefore we allow them to contain 
non-renormalizable interactions. First, we examine 
the case where the goldstino is a gauge singlet. We 
elucidate the model-independent relations between
supersymmetry-breaking masses and some associated 
interactions, with special attention to the behaviour 
of some four-particle scattering amplitudes in the high- 
and low-energy limits. The former gives constraints 
from tree-level unitarity, the latter affects the 
phenomenological lower bounds on the gravitino mass.
In particular, we give new results for the annihilation
of photons into gravitinos, and discuss their implications. 
We then examine the case with no neutral chiral 
superfields, so that the goldstino is charged, and the 
gauge symmetry is also broken. In this context, we
discuss the singularity structure and the associated 
unitarity constraints, relating the scales of supersymmetry
and gauge symmetry breaking. We conclude by commenting on 
possible realistic examples, where the broken gauge symmetry 
is associated with grand or electroweak unification.
\end{abstract}
\vfill{
CERN-TH/96-369
\newline
\noindent
December 1996}
\end{titlepage}
\setcounter{footnote}{0}
\vskip2truecm
\nsect{Introduction}

According to the present theoretical wisdom, space-time
supersymmetry is likely to play a r\^ole in the unification
of all fundamental interactions, including gravity, and in
the resolution of the naturalness problem of the Standard
Model (for reviews and references, see e.g. \cite{susyrev}).
The major obstacle to the construction of a predictive
supersymmetric extension of the Standard Model is our
limited understanding of the physics underlying the
spontaneous breaking of supersymmetry. However, the general
`kinematical' aspects of the problem are well understood
\cite{global,local}. If $\lam_S$ is the scale of supersymmetry
breaking, the mass splittings in the different sectors of the
model are given by $\Delta m^2 \sim  \lambda \, \lsu^2$, where
$\lambda$ is the effective coupling of the goldstino supermultiplet
(whose fermionic degrees of freedom provide the $\pm 1/2$ helicity
components of the massive spin-$3/2$ gravitino) to the sector under
consideration. Even assuming a given size for the mass splittings
(for example, the electroweak scale), this leaves an ambiguity in 
the size of $\lam_S$. If $\lam_S^4 < \Delta m^2 \, \mpl^2$, where
$\mpl \equiv (8 \pi G_N)^{-1/2} \simeq 2.4 \times 10^{18} \gev$
is the (reduced) Planck mass, the effective couplings of the goldstino
supermultiplet to the other chiral and vector supermultiplets are
stronger than the gravitational ones. Therefore, for most practical
purposes, we can discuss spontaneous supersymmetry breaking in an 
effective non-renormalizable theory with global supersymmetry: 
formally, we can start from the general supergravity formalism 
\cite{sugraym} and take the limit $\mpl \to \infty$, while keeping 
$\lam_S$ constant \cite{faylim}. We just need to recall that
in the physical spectrum there is a gravitino of mass $m_{3/2} = 
\Lambda_S^2 / (\sqrt{3} \mpl)$, whose interactions at energies 
$E \gg m_{3/2}$ are well approximated, thanks to an equivalence 
theorem \cite{susyeq}, by those of the massless goldstino in the
global limit. On physical grounds, models with a relatively low 
scale of supersymmetry breaking may be favoured by arguments 
related with cosmology and with the flavour problem, which explains 
their recent revival \cite{recent}.

In the study of low-energy supersymmetry breaking, two different and
complementary approaches can be followed. The `microscopic' approach
tries to understand the symmetries and the dynamics that may generate
the scales $\lam_S$ and $\Delta m^2$ at the level of the fundamental
theory. The `macroscopic' approach tries to write down an effective
theory, valid in a suitable energy range, where the underlying
dynamics is encoded in a set of non-renormalizable interactions, and 
the spontaneous breaking of supersymmetry is described at the 
classical level. Intermediate approaches are possible \cite{recent}, 
but we will not discuss them here. The `macroscopic' approach has a 
limited predictive power, but may be useful to parametrize the 
model-independent features of the resulting phenomenology, modding 
out the dependence on the presently unknown aspects of the 
fundamental theory. In this paper, we adopt such an approach, with 
the purpose of discussing some aspects of spontaneous supersymmetry 
breaking in the presence of gauge interactions.

In section~2, we examine the case where the goldstino is a gauge
singlet and the gauge symmetry remains unbroken. As an 
illustrative example, we study a class of models with gauge group 
$U(1)$ and two neutral chiral superfields, $S$ and $A$, which is 
sufficient to understand most qualitative aspects of the general 
case. In particular, we consider the model-independent relations 
between various supersymmetry-breaking masses and some associated 
interactions. Many results of this section are not new, but we try 
to give a simple and systematic presentation, avoiding the 
unnecessary use of the supergravity formalism, and stressing the 
general cancellation mechanisms that
control the low- and high-energy behaviour of some four-particle 
scattering amplitudes. In the low-energy limit, we extract some
useful information on the phenomenology of a superlight gravitino
within realistic models. In particular, we clarify the structure 
of the low-energy interactions between photons and gravitinos, which 
was used in the past \cite{mmy,ghe,gmr} to establish lower bounds on 
the gravitino mass, via the r\^ole played by processes such as 
photon--photon annihilation into gravitino pairs (or vice versa) 
in cosmological nucleosynthesis or in the evolution of supernovae.
To do so, we generalize previous calculations of the relevant
scattering amplitudes and cross-sections. In the high-energy
limit, we use tree-level unitarity to put an upper bound on the
energy range of validity of the effective theory, as a function
of the particle masses and of the scale of supersymmetry breaking.
We conclude the section with simple examples of explicit models.

In section~3, we examine models that do not contain neutral
superfields, so that the gauge symmetry must be broken 
simultaneously with supersymmetry. We construct a minimal example, 
with gauge group $U(1)$ and two chiral superfields $(H_1,H_2)$ of 
opposite charges, which allows us to discuss in simple terms a 
number of general properties of possible physical relevance. We 
show that, in order to have a supersymmetry-breaking vacuum, either 
the superpotential or the K\"ahler metric must be singular at the 
origin, so that the configurations with unbroken gauge
group are excluded from the space of allowed background field values.
Considering first the simple case of canonical kinetic terms, we
show that the unique choice of superpotential leading to a
supersymmetry-breaking classical vacuum has a conical singularity 
at the origin, $w = \lam_S^2 \sqrt{2 H_1 H_2}$ (apart from 
irrelevant constant terms and redefinitions of $\lam_S$). We also 
show how this model can be obtained from a supergravity model with 
vanishing vacuum energy, previously studied in ref.~\cite{bfz}, by 
taking the appropriate flat limit. We then discuss how the scales
$\lsu$ and $v$, associated with supersymmetry and gauge symmetry 
breaking, respectively, control the spectrum and the interactions, 
and show that $\lsu \gg v$ would lead to a premature violation of 
tree-level unitarity and correspond to a strong coupling r\'egime. 
We continue by extending the treatment to the case of non-canonical 
kinetic terms, with the formulation of an explicit example of this 
sort. We conclude by generalizing our results to a much wider class of
models, and commenting on possible realistic applications, 
where supersymmetry breaking is associated with the
breaking of either a grand-unified symmetry or the
electroweak symmetry.

Some notational and calculational details are confined to the
appendix, which contains the explicit form of an $N=1$ globally
supersymmetric effective Lagrangian, in the simple case where
the gauge group is a non-anomalous $U(1)$ and there is no
Fayet--Iliopoulos term, as well as the corresponding expression 
for the scalar potential in the locally supersymmetric case.
\nsect{The case of a singlet goldstino}

As anticipated in the introduction, we will examine in this section,
in an effective theory approach, some model-independent
features of $N=1$ globally supersymmetric gauge theories coupled
to matter, with spontaneously broken supersymmetry and the goldstino
transforming as a gauge singlet. Our aim is to make explicit some 
relations, associated with low-energy theorems \cite{let}, between 
supersymmetry-breaking masses and certain cubic and quartic 
interactions involving the goldstino and/or its spin-0 superpartners. 
These interactions in turn control the low- and high-energy behaviour 
of some four-particle scattering amplitudes, which can be used to 
establish lower bounds on the scale of supersymmetry breaking (and 
therefore on the gravitino mass) and to constrain the range of 
validity of the effective theory in energy and parameter space. 

For our purpose, it will be sufficient to consider a class of $N=1$
globally supersymmetric models\footnote{For the notation, we
refer to the appendix.} with gauge group $G=U(1)$ and two neutral
chiral superfields, with physical components $(S,{\tilde S})$ and
$(A,{\tilde A})$. We anticipate that the two chiral superfields
will play quite different r\^oles in the following. The $S$-multiplet
will be directly responsible for supersymmetry breaking and will
contain the goldstino. Indeed, most of the considerations of the
present section do not require the presence of the $A$-multiplet.
We introduce it here for two reasons: first, to develop a formalism
that can be easily adapted to the generalization presented in the 
next section; second, because $A$ mimics the r\^ole of matter 
superfields in realistic models. We recall that the most general 
effective Lagrangian\footnote{We remind the reader that this
formulation is the most general one up to terms with at most 
two derivatives in the bosonic fields: higher-derivative terms
are not included.} with the above field content is determined by
a superpotential $w$, a K\"ahler potential $K$ and a gauge kinetic 
function $f$. Incidentally, notice that, since the chiral superfields 
are neutral, the gauge kinetic function is the only source of 
interactions between vector and chiral multiplets. Having in mind the 
different r\^oles to be played by $S$ and $A$, we will retain the full 
$S$-field dependence of the basic functions $w,K,f$ and consider only 
a power expansion in the $A$ field. More precisely, we will assume the 
following functional forms:
\bea
\label{wappc}
w & = & \wh (S) + {1 \over 2} \wt(S) A^2 + \ldots \, ,
\\
K & = & \kh (S,\ov{S}) + \kt (S,\ov{S}) |A|^2 + {1 \over 2}
[ \htil (S,\ov{S}) A^2 + {\rm h.c.} ] + \ldots \, ,
\label{kappc}
\\
f & = & \fh (S) + {1 \over 2} \ft (S) A^2 + \ldots \, ,
\label{fappc}
\eea
where the dots stand for higher-order terms in $A$. Since there 
are no charged fields, the classical potential $V$ is just $||F||^2
=w_i (K^{-1})^{i \, {\ov \j}} {\ov w}_{\ov \j}$, as in eq.~(A.11). 
We assume that $w$ and $K$ are such that, at the minimum of $V$:
\be
\label{cond}
\langle F^S \rangle \ne 0 \, ,
\;\;\;\;\;
\langle F^A \rangle  =  0 \, ,
\;\;\;\;\;
\langle   A \rangle  =  0  \, ,
\ee
whereas we do not need to specify whether $\langle S \rangle$
is vanishing or not. Also, the fact that $\langle A \rangle = 0$ and 
$\langle F^A \rangle = 0$ are at least local extrema is already a 
consequence of eqs.~(\ref{wappc}) and (\ref{kappc}).
Indeed, only the assumption that $\langle F^S \rangle \ne 0$
is restrictive, since it amounts to requiring that supersymmetry
be spontaneously broken. For the purposes of this section, however,
we do not need to commit ourselves to a specific choice of $\wh (S)$
and $\kh (S,\ov{S})$ that realizes such a situation.

\subsection{The spectrum}

We begin the study of our class of models by discussing the general 
features of the mass spectrum. Starting with the supersymmetry-breaking
sector, we can identify the goldstino with the fermion $\tilde{S}$: 
$m_{\tilde{S}}=0$. We can verify this with the help of eq.~(A.17) and 
of the minimization condition for the scalar potential along the $S$ 
direction:
\be
\langle 
( \log \wh_S )_S
- 
( \log \kh_{{\ov S} S} )_S
\rangle 
= 0 \, .
\label{minis}
\ee
Decomposing the shifted $S$ field into real and imaginary parts,
\be
S \equiv \langle S \rangle + {\varphi_S + i \varphi'_S \over \sqrt{2}}
\, ,
\label{reims}
\ee
and assuming for simplicity that $\langle V_{S S} \rangle = \langle 
V_{\ov{S} \ov{S}} \rangle$ (the generalization is straightforward),
$\varphi_S$ and $\varphi_{S'}$ are mass eigenstates, with eigenvalues
\bea
m^2_S &  = & \langle |F^S|^2
[ - ( \log \kh_{{\ov S} S} )_{{\ov S} S}
  - ( \log \kh_{{\ov S} S} )_{S S}
  + ( \log \wh_S )_{S S} ] \rangle \, ,
\nonumber \\
m^2_{S'} & = & \langle |F^S|^2
[ - ( \log \kh_{{\ov S} S} )_{{\ov S} S}
  + ( \log \kh_{{\ov S} S} )_{S S}
  - ( \log \wh_S )_{S S} ] \rangle \, ,
\label{smasses}
\eea
where $F^S = (K^{-1})^{S \ov{\j}} \ov{w}_{\ov{\j}}$. 
The sum of the squared scalar masses in the $S$-sector
has a geometrical meaning \cite{fgp}, and is given by
\be
m^2_S + m^2_{S'} = 2 \langle |F^S|^2
[ - ( \log \kh_{{\ov S} S} )_{{\ov S} S} ]
\rangle  = - 2 \left\langle |F^S|^2 {R_{S {\ov S} S {\ov S}}
\over  \kh_{{\ov S} S}} \right\rangle \, ,
\label{ssrr}
\ee
where $R_{S \ov{S} S \ov{S}} \equiv \kh_{S \ov{S} S \ov{S}}
- |\kh_{\ov{S} S S}|^2 / \kh_{\ov{S} S}$ is the curvature
of the K\"ahler manifold for the complex field $S$. Notice
that, in order to have a stable minimum in the $S$ direction, one
needs $m_S^2 + m_{S'}^2 > 0$, therefore $R_{S \ov{S} S \ov{S}} <0$.
This can be contrasted with hidden sector breaking in supergravity 
models, where one usually deals with manifolds of positive curvature,
so that curvature contributions to scalar squared masses are negative.
In such a case, however, positivity can be rescued by an additional
non-negligible positive contribution to scalar squared masses coming 
from gravitational interactions and proportional to the gravitino mass.

Moving to the gauge sector, the $U(1)$ gauge boson is massless,
since the gauge symmetry is unbroken: $m_V = 0$. The gaugino is
a mass eigenstate, with mass scaling with $F^S$ and controlled
by the derivative of $\fh$:
\be
M = {1 \over 2} \left\langle {\fh_S F^S \over {\rm Re \,} \fh} 
\right\rangle \, .
\label{gaum}
\ee
We recall that gaugino masses have expressions of this type
also in supergravity models.

Notice that, if the kinetic terms were canonical, there would be no
supersymmetry-breaking mass splittings in the $S$ and in the
gauge sectors. This is a consequence of a general result of
global supersymmetry \cite{fgp}, which can be conveniently described 
in terms of ${\rm Str \;}{\cal M}^{2} \equiv \sum_i (-1)^{2 J_i}
(2 J_i + 1) m_i^2$, where $m_i$ and $J_i$ are mass and spin of the i-th
physical state: for canonical kinetic terms
and in the absence of anomalous $U(1)$ gauge factors, $\str
{\cal M}^2 \equiv 0$ in each separate sector of the mass
spectrum. In the $(S,\tilde{S})$ sector, which contains the
massless goldstino, no mass splitting would be allowed. In
the $(\lambda,A_{\mu})$ sector, which contains a massless
gauge boson, the gaugino would also be massless. As we have
seen, this trivial pattern is changed by the introduction
of non-canonical kinetic terms.

Non-canonical kinetic terms play an important r\^ole also 
in the spectrum of the $A$ sector. Here the fermion mass is
\be
m_{\tilde{A}} =
\langle \kt^{-1} ( \wt - \ov{F}^{\ov{S}} \htil_{\ov{S}} )
\rangle \, .
\label{mferm}
\ee
The elements of the spin-0 mass matrix are, in obvious notation:
\be
m^2_{A \ov{A}} =
|m_{\tilde{A}}|^2 + \tilde{m}^2_A \, ,
\label{mdiag}
\ee
where 
\be
\tilde{m}^2_A =  \langle
\ov{F}^{\ov{S}} ( - \log \kt )_{\ov{S} S} F^S
\rangle \, ,
\label{softm}
\ee
and
\be
m^2_{A A} = \left\langle
[ \kt^{-1} \wt ] \left[
\left( \log {\wt \over \kt^2} \right)_S
F^S \right]
+
[- \kt^{-1} \ov{F}^{\ov{S}} \htil_{\ov{S}} ]
\left[ \left( \log {\htil_{\ov{S}} \over \kt^2}
\right)_S F^S \right] \right\rangle \, .
\label{moffd}
\ee
Under the simplifying assumption that $m^2_{A A}$ is real
(the generalization is straightforward), the real and 
imaginary parts of the field $A$, 
\be
A \equiv {\varphi_A + i \varphi'_A  \over \sqrt{2}} \, ,
\label{reima}
\ee
are mass eigenstates, and the corresponding eigenvalues are 
\be
m_A^2 = m_{A \ov{A}}^2 + m^2_{AA} \, ,
\;\;\;\;\;
m_{A'}^2 = m_{A \ov{A}}^2 - m^2_{AA} \, .
\label{amasses}
\ee
Notice that also the masses of the $A$ sector have some formal similarity 
with those derived in supergravity models. The fermion mass $m_{\tilde A}$ 
is the sum of two pieces (`generalized $\mu$ parameters'), one 
coming directly from the superpotential, the other induced from the 
K\"ahler potential and scaling with $F^S$. The diagonal scalar 
mass $m^2_{A \ov{A}}$ is the sum of the squared fermion 
mass and an additional contribution $\tilde{m}^2_A$ 
(`diagonal soft mass'), scaling with $|F^S|^2$ and controlled by the 
curvature of the $S$--$A$ K\"ahler manifold. The off-diagonal scalar 
mass $m^2_{A A}$ is the sum of two pieces, proportional 
to the corresponding pieces composing the fermion mass, with 
proportionality factors scaling with $F^S$ (`generalized $B$ 
parameters').

\subsection{Cubic interactions}

The non-canonical forms of $K$ and $f$, which are at the origin
of the supersymmetry-breaking mass splittings in the various
sectors, also generate non-trivial interactions, which we will
now discuss in a systematic way, applying the general formulae
of the appendix. Here and in the following, when writing down
interaction terms derived from expansions around a given vacuum,
we will work with canonically normalized fields, without introducing
a different symbol to denote them. We will also introduce the quantity 
$F \equiv \langle {\ov w}_{\ov S} (K_{\ov S S})^{-1/2} 
\rangle$, with the property that $|F|^2 = \langle ||F||^2 \rangle$.
Whenever possible, we will try to express the couplings in terms
of the physical masses and of the supersymmetry-breaking 
scale\footnote{Both in the introduction and in the explicit
models presented below, the supersymmetry-breaking scale
is denoted also by the symbol $\lsu$. We anticipate that, in the 
explicit models, the definition of $\lsu$ will be such that 
$\lsu^4=\langle ||F||^2 \rangle=|F|^2$.} $F$.
In this subsection we will focus on cubic interactions
and some related decay processes. We will also impose a sort of
minimal unitarity (or perturbativity) requirement, by asking 
that decay widths be not larger than the masses of the decaying 
particles. This will lead to simple relations where (combinations of) 
masses are bounded from above by the supersymmetry breaking scale.
Also, we anticipate that the cubic couplings studied here  
play a crucial r\^ole in building the four-particle amplitudes
discussed in the next subsection. 

\subsubsection{$S \gamma \gamma$}

We start by expanding the generalized kinetic terms for the gauge boson,
eq.~(A.8), which leads to the following cubic couplings involving
two photons and the spin-0 components of $S$:
\be
\label{t1a}
-{1 \over 2 \sqrt{2}} \left[ {\rm Re \,}  \left( M \over F
\right) \varphi_S - {\rm Im \,}  \left( M \over F
\right) \varphi_S' \right] F_{\mu \nu} F^{\mu \nu}
\ee
\be
\label{t1b}
+ {1 \over 4 \sqrt{2}} \left[ {\rm Im \,}  \left( M \over F
\right) \varphi_S + {\rm Re \,}  \left( M \over F
\right) \varphi_S' \right]
\epsilon^{\mu \nu \rho \sigma} F_{\mu \nu} F_{\rho \sigma}
\, .
\ee
The physical processes directly controlled by these couplings
are the decays $\varphi \rightarrow \gamma \gamma$, where
$\varphi$ denotes any of the two fields $\varphi_S$ and
$\varphi_S'$, and $\gamma$ denotes the massless gauge boson 
(`photon'). The decay rate is easily computed to be:
\be
\Gamma(\varphi \to \gamma \gamma) =
{m_{\varphi}^3 \over 32 \pi} {|M|^2 \over |F|^2} \, ;
\ee
then the requirement $\Gamma_{\varphi} < m_{\varphi}$ translates 
into the following bound:
\be
\label{bsgf}
m_{\varphi} | M | <  \sqrt{32 \pi} |F| \, .
\ee

\subsubsection{$S \lambda \lambda$}

Expanding the gaugino bilinears, eqs.~(A.13) and
(A.16), and making use of the gaugino equation of
motion, we get the following effective Yukawa couplings
involving two gauginos and the spin-0 components of $S$:
\be
\label{t2}
{1 \over 4 \sqrt{2} } 
\left[ - {4 M^2 \over F}
+  \left\langle  { \fh_{SS} - \fh_S  \Gamma^{S}_{SS} \over   
{\rm Re \,} \fh \, \kh_{\ov{S} S} } \right\rangle  F
\right] ( \varphi_S + i \varphi_S') \lambda 
\lambda + {\rm h.c.} \, ,
\ee
where $\Gamma^{S}_{SS}  \equiv (K^{-1})^{S \ov{\i}} 
K_{\ov{\i} S S}$ and $\langle \Gamma^{S}_{SS} \rangle 
= \kh_{\ov{S} S S} /  \kh_{\ov{S} S}$. Of course, the 
gauginos appearing in these effective couplings
are understood to be external, in a physical process, 
otherwise the original derivative vertex should be used. 
A possible application is a decay of the type
$\varphi \to \lambda \lambda$, which is kinematically 
allowed only if $|m_{\varphi}| > 2 |M|$. 

\subsubsection{$S \tilde{S} \tilde{S}$}

Expanding the bilinears in the goldstino $\tilde{S}$,
eqs.~(A.15) and (A.17), and making use of the
goldstino equation of motion, we get the following effective 
Yukawa couplings involving two goldstinos and the spin-0 
components of $S$, in agreement with \cite{ferru}:
\be
\label{t3}
- {1 \over 2 \sqrt{2}}
{m_S^2 \over F}  \varphi_S \tilde{S} \tilde{S}
+
{i \over 2 \sqrt{2}}
{m_{S'}^2 \over F}  \varphi_S' \tilde{S} \tilde{S}
+ {\rm h.c.} 
\ee
The physical processes directly controlled by these couplings
are the decays $\varphi \rightarrow \tilde{S} \tilde{S}$. The
decay rate is easily computed to be:
\be
\label{gsss}
\Gamma(\varphi \to \tilde{S} \tilde{S}) =
{m_{\varphi}^5  \over 32 \pi |F|^2} \, ;
\ee
then the requirement $\Gamma_{\varphi} < m_{\varphi}$ translates 
into the following bound:
\be
\label{bsss}
m_{\varphi}^2 <   \sqrt{32 \pi} |F| \, .
\ee

\subsubsection{$S S S$}

Expanding the scalar potential, eq.~(A.10), one can
generate cubic couplings in the spin-0 components of $S$,
controlled by:
\be
\langle V_{\ov{S}SS} \rangle = |F|^2 \left\langle
- \left( \log \kh_{\ov{S} S} \right)_{\ov{S}SS}
\right\rangle \, ,
\ee
\be
\langle V_{SSS} \rangle = |F|^2 \left\langle
- \left( \log \kh_{\ov{S} S} \right)_{SSS}
+ \left( \log \wh_S \right)_{SSS}
\right\rangle \, .
\ee
As examples of physical processes controlled by the above couplings, one
could think of the possible decays $\varphi \to \varphi'
\varphi'$, if they are kinematically allowed.

\subsubsection{$\tilde{S} \lambda \gamma$}

Expanding the gaugino--goldstino bilinear, eq.~(A.14),
one generates a cubic photino--goldstino--photon coupling of
the form
\be
\label{dipole}
- {1 \over \sqrt{2}} \displaystyle{M \over F} \tilde{S}
\sigma^{\mu \nu} \lambda F_{\mu \nu} + {\rm h.c.} 
\ee
The physical process directly controlled by this coupling
is the decay $\lambda \rightarrow \gamma \tilde{S}$. The
decay rate is easily computed to be \cite{fayetph}:
\be
\label{gdip}
\Gamma(\lambda \to \gamma \tilde{S}) =
{|M|^5  \over 16 \pi |F|^2} \, ;
\ee
then the requirement $\Gamma_{\lambda} < |M|$ translates into 
the following bound:
\be
\label{bdip}
|M|^2 <  \sqrt{16 \pi} |F| \, .
\ee

\subsubsection{$S \tilde{A} \tilde{A}$, $A \tilde{A} \tilde{S}$}

Finally, expanding eqs. (A.15) and (A.17) and using again
the fermion equations of motion, we get the following
effective Yukawa interactions coupling the $S$-sector to 
the $A$-sector:
\be
\label{t6a}
- {1 \over 2 \sqrt{2}} \left\{ {m_{A}^2 - m_{A'}^2 \over 
2 F} - \langle { (\htil_{SS} - \htil_S \Gamma^S_{SS})^*
\over \kt \kh_{\ov{S} S} } \rangle F^* \right\} \varphi_S 
\tilde{A} \tilde{A} + {\rm h.c.} \, ,
\ee
\be
\label{t6b}
+ {i \over 2 \sqrt{2}} \left\{ - {m_{A}^2 - m_{A'}^2 \over 
2 F} - \langle { (\htil_{SS} - \htil_S \Gamma^S_{SS})^* 
\over \kt \kh_{\ov{S}S}} \rangle F^* \right\} \varphi_S' 
\tilde{A} \tilde{A} + {\rm h.c.} \, ,
\ee
\be
\label{t7}
- {m_A^2 - m_{\tilde{A}}^2 \over \sqrt{2} F}
\varphi_A \tilde{A} \tilde{S}
+ i
{m_{A'}^2 - m_{\tilde{A}}^2 \over \sqrt{2} F}
\varphi'_A \tilde{A} \tilde{S}
+ {\rm h.c.}
\ee
The latter couplings involving the goldstino are especially simple
and are manifestly related to the mass splittings in the 
$(A,{\tilde A})$ multiplet, similarly to the cases considered above.
The decay rates for the associated processes 
$\varphi_A \rightarrow {\tilde A} {\tilde S}$
(or ${\tilde A} \rightarrow \varphi_A {\tilde S}$) and
$\varphi'_A \rightarrow {\tilde A} {\tilde S}$
(or ${\tilde A} \rightarrow \varphi'_A {\tilde S}$) are easily 
evaluated. The general result for decays of this kind is
\be
\Gamma =
{(\Delta m^2)^4 \over 16 \pi m^3 |F|^2} \, ,
\ee
where $\Delta m^2$ denotes the mass splitting and $m$ the
mass of the decaying particle. The requirement $\Gamma < m$ 
now translates into bounds of the form
\be
\label{sabound}
{ (\Delta m^2)^2 \over m^2 } <  \sqrt{16 \pi} |F| \, .
\ee
In particular, if the final particle produced together with the 
goldstino is massless, then $\Delta m^2=m^2$ and the above results 
simplify to $\Gamma = m^5/(16 \pi |F|^2)$ and $m^2 < \sqrt{16 \pi} |F|$,
which are the same as (\ref{gdip}) and (\ref{bdip}) for
gauginos. The analogous relations (\ref{gsss}) and (\ref{bsss}) 
for the scalar partners of the goldstino have an extra factor 
$1/2$ just because of the identical particles in the final state.

The bounds of eqs.~(\ref{bsgf}), (\ref{bsss}), (\ref{bdip})
and (\ref{sabound}) show that the typical supersymmetry-breaking
mass splittings should not be larger than a few times the 
supersymmetry-breaking scale, otherwise some coupling constant
would become large and a particle interpretation would become
problematic for some of the states explicitly considered here.  

\subsection{Quartic interactions}

We have seen how the presence of non-trivial functions 
$w$, $K$ and $f$ generates mass terms and cubic interactions.  
Several quartic and higher interactions are generated as well.
For instance, the quartic interactions can involve four bosons, 
two bosons and two fermions, or four fermions. Instead of discussing
all of them, we will mainly focus on four-particle physical amplitudes 
involving goldstinos, which are more model-independent and exhibit 
peculiar low- and high-energy properties. Such amplitudes may receive 
both `contact' contributions, directly from the Lagrangian, and 
`exchange' contributions, where two cubic interactions are connected 
by a particle propagator. In the low-energy r\'egime, important 
cancellations make the amplitudes vanish with energy more rapidly 
than the individual contributions. This softer behaviour is consistent 
with the derivative character of the effective goldstino interactions 
at low energy. In the high-energy r\'egime, the contact terms tend 
to dominate and make the amplitudes grow until unitarity 
breaks down. In this r\'egime, the requirement of tree-level unitarity 
can be used to get upper bounds on the energy itself, i.e. on 
the validity of the effective theory, as a function of the particle 
masses and of the supersymmetry breaking scale. Notice that,
in the examples discussed below, the source of unitarity violations
is the non-canonical character of $K$ and $f$: the relevant contact
terms would be absent in the case of canonical kinetic terms.
We also recall that, in the framework of an underlying supergravity theory, 
an equivalence theorem \cite{susyeq} assures that the goldstino 
components of the gravitino give the leading contribution 
to gravitino scattering amplitudes when $s \gg m^2_{3/2}$. 
This condition will always be understood in the following, 
even when we discuss the low-energy r\'egimes $s\ll ($masses$)^2$.
In the high-energy r\'egime, our formalism will lead to considerable
simplifications with respect to the full supergravity formalism,
since the cancellation mechanisms, implied by the equivalence 
theorem, will be automatically taken into account from the beginning.

\subsubsection{$\tilde{S} \tilde{S} \tilde{S} \tilde{S}$}

The simplest amplitudes that exhibit the above properties are those 
involving four goldstinos, a case already discussed in \cite{ferru}
in connection with the goldstino--gravitino equivalence theorem: the
relevant Feynman diagrams are shown schematically in fig.~1.
\begin{figure}
\centerline{
\epsfig{figure=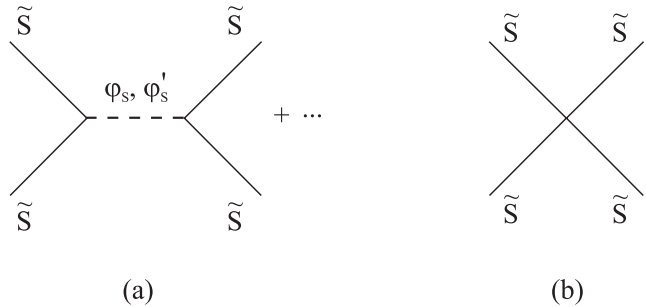,height=4cm,angle=0}
}
\caption{\it Tree--level Feynman diagrams contributing to the
four-goldstino scattering amplitudes: (a) $(\varphi_S, \varphi_S')$
exchange (the dots stand for the $t$- and $u$- channel diagrams);
(b) contact interactions.}
\label{figuno}
\end{figure}
As an illustrative example, let us reconsider the scattering amplitude 
${\cal A}={\cal A} ({\tilde S}_- {\tilde S}_- \rightarrow {\tilde S}_- 
{\tilde S}_-)$, involving four left-handed goldstinos ${\tilde S}_-$. 
This amplitude receives two contributions from the $s$-channel exchange 
of the scalars ${\varphi_S}$ and ${\varphi_S'}$ (fig.~1a), which couple 
to the external goldstinos through the cubic vertices (\ref{t3}). In 
addition, ${\cal A}$ receives a contact contribution (fig.~1b) from the 
four-fermion Lagrangian term  
\be
\label{four}
- {1 \over 8}
{ m_S^2 + m_{S'}^2 \over |F|^2 } \tilde{S} \tilde{S}
\ov{\tilde{S}} \ov{\tilde{S}} \, ,
\ee
which originates from eqs.~(A.23) and (\ref{ssrr}). The three 
contributions give
\bea
{\cal A} & = & {s \over 2 |F|^2} \left( {m_S^4 \over s - m_S^2}  
+ {m_{S'}^4 \over s - m_{S'}^2} + (m_S^2+m_{S'}^2) \right) 
\label{asum} 
\\
& = & {s^2 \over 2 |F|^2} \left( {m_S^2 \over s - m_S^2}  
+ {m_{S'}^2 \over s - m_{S'}^2} \right) \, , 
\eea
where an overall $s$ factor is just due to the external spinors.
In the low-energy limit ($s \ll m_S^2, m_{S'}^2$), one can clearly
see the cancellation of the leading individual terms, proportional to 
$s m^2/|F|^2$. The resulting amplitude $|{\cal A}| = s^2/|F|^2$
has a softer energy dependence, does not depend on the scalar 
masses, and gives a cross section $\sigma = s^3/(32 \pi |F|^4)$. 
At high energies ($s > m_S^2, m_{S'}^2$), on the other hand, the 
contact term alone dominates and produces an amplitude 
scaling as $|{\cal A}| \sim s m^2/|F|^2$, which signals a unitarity 
breakdown for $s \gg |F|^2/m^2$. The corresponding cross
section grows as $\sigma \sim s m^4 /|F|^4$. 
An accurate determination of the critical energy can be
inferred from the detailed analysis of \cite{ferru}, where 
all four-goldstino helicity amplitudes were computed
and the whole scattering matrix was built. Requiring that
the largest eigenvalue of the matrix be smaller than 1 in the
high energy r\'egime implies that unitarity is violated beyond 
the critical energy
\be
s_c=48\pi\frac{\vert F \vert^2}{m_S^2+m_{S'}^2} \, .
\label{scri}
\ee
Therefore the description provided by the model is not reliable 
beyond this energy cutoff, and we should require $s_c$ to be larger
than the scalar masses for consistency reasons.
If we impose, for example, $s_c>m_S^2+m_{S'}^2$ we get the bound
\be
m_S^2+m_{S'}^2 < \sqrt{48 \pi} |F| \, ,
\label{bscri}
\ee
which has the same form as (and is consistent with) the previous
bound (\ref{bsss}).

\subsubsection{$\tilde{S} \tilde{S} \gamma \gamma$}

Another class of interesting processes consists of those
involving two goldstinos and two gauge bosons (`photons'),
i.e. the annihilation processes ${\tilde S} {\tilde S} 
\rightarrow \gamma \gamma$ and $\gamma \gamma \rightarrow 
{\tilde S} {\tilde S}$, and the scattering process $ {\tilde S}
\gamma \rightarrow {\tilde S} \gamma$, described by the Feynman 
diagrams of fig.~2. 
\begin{figure}
\centerline{
\epsfig{figure=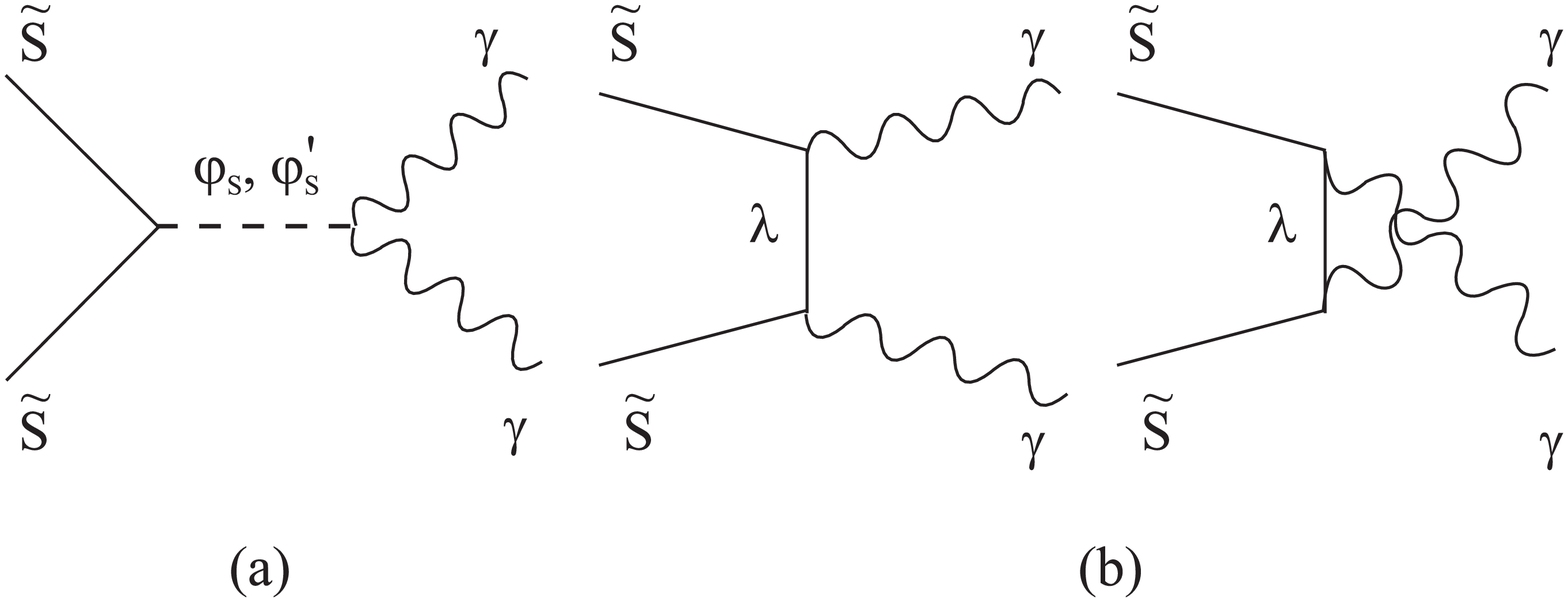,height=4cm,angle=0}
}
\caption{\it Tree--level Feynman diagrams contributing to the
scattering amplitudes with two photons and two goldstinos, for
example $\tilde{S} \tilde{S} \to \gamma \gamma$: (a) $(\varphi_S, 
\varphi_S')$ exchange; (b) gaugino exchange.}
\label{figdue}
\end{figure}
The Lagrangian does not contain any contact 
term contributing to these processes. On the other hand, one has 
to take into account contributions from virtual ${\varphi_S}, 
{\varphi_S'}$ exchange (fig.~2a), controlled by the cubic couplings 
(\ref{t1a}), (\ref{t1b}), (\ref{t3}), and virtual gaugino exchange
(fig.~2b), controlled by the cubic coupling (\ref{dipole}). We first 
consider the low-energy r\'egime ($s \ll m_S^2, m_{S'}^2,|M|^2$), where 
again the full amplitudes are softer than the individual contributions,
owing to a crucial cancellation. A straightforward way to check such a 
cancellation consists in integrating out the massive fields from the 
Lagrangian and looking at the leading effective Lagrangian 
terms coupling two goldstinos and two gauge bosons. Integrating out the 
scalars leads to an effective term
\be
{ M \over 8 F^2} \left( F_{\mu\nu} F^{\mu\nu} + {i \over 2} 
\epsilon^{\mu\nu\rho\sigma} F_{\mu\nu} F_{\rho\sigma} \right) 
{\tilde S}{\tilde S} + {\rm h.c.} \, ,  
\ee
which gives contributions $\sim E^3 M/F^2$ to the amplitudes, whereas
integrating out the gauginos leads to an identical effective term 
with opposite sign. Therefore the amplitudes do not receive any net 
contribution proportional to $E^3$. The next term in the expansion
couples ${\tilde S}$ to ${\ov {\tilde S}}$ and contains an extra 
derivative, coming from the gaugino propagator:
\be
- {i \over 2 |F|^2} F_{\mu\nu} {\ov {\tilde S}} 
{\ov {\sigma}}^{\mu \nu} {\ov {\sigma}}^{\tau} 
\sigma^{\rho \sigma} \partial_{\tau} ( \tilde{S}
F_{\rho \sigma} ) 
= -{i \over 2|F|^2} F_{\mu\nu} F^{\mu}_{\;\; \tau}
\left[ {\ov{\tilde S}} {\ov \sigma}^{\nu} \partial^{\tau}{\tilde S}
- (\partial^{\tau} {\ov{\tilde S}}) {\ov \sigma}^{\nu} {\tilde S} \right]
\, ,
\label{effop}
\ee
where the equations of motion for the goldstino and the photon have 
been used in the second step. The amplitudes corresponding to the 
above effective operator are proportional to $E^4/|F|^2$ and do not 
depend on masses. This behaviour is analogous to the one of the
four-goldstino amplitude discussed above and leads to cross sections 
scaling again as $\sigma \sim s^3/|F|^4$. The important cancellation explained
above was not accounted for by the recent calculation of ref.~\cite{ghe}, 
where a result $\sigma \sim s^2 |M|^2/|F|^4$ was obtained for the 
annihilation process ${\tilde S} {\tilde S} \rightarrow \gamma \gamma$.
Indeed, existing calculations are only valid either in the high-energy
r\'egime \cite{bruno} or in the massless limit for the spin-0
partners of the goldstino \cite{brdue,ghe}. More details and a
discussion of the phenomenological implications of our results
will be given in section~2.4.
We now consider the above processes in the high-energy r\'egime 
($s > m_S^2, m_{S'}^2,|M|^2$), where unitarity violations are 
expected. Here the most dangerous amplitudes correspond to external 
${\tilde S}$, ${\ov {\tilde S}}$ and are dominated by gaugino exchange. 
They grow as $E^2 |M|^2/|F|^2$, which signals a unitarity breakdown 
for $s \gg |F|^2/|M|^2$. The corresponding cross sections grow as 
$\sigma \sim s |M|^4 /|F|^4$. Imposing the consistency condition that 
the critical energy be larger than $|M|$ leads to a bound 
of the same form as the previous bound (\ref{bdip}), $|M|^2 < {\cal O} 
(10) |F|$. Similarly, imposing the consistency condition that the
critical energy be larger than $m_{\varphi}$, where $\varphi$ is now
the heavier of $\varphi_S$ and $\varphi_{S'}$, leads to a bound 
of the same form as the previous bound (\ref{bsgf}), $| m_{\varphi} 
M| < {\cal O} (10) |F|$. Both bounds are in agreement with the
supergravity results of ref.~\cite{brdue}.

\subsubsection{$\tilde{S} \tilde{S} \lambda \lambda$}

We now move to processes involving two goldstinos and two gauginos,
i.e. $ \tilde{S} \tilde{S} \to \lambda \lambda $, $\lambda \lambda 
\to \tilde{S} \tilde{S}$ and $\tilde{S} \lambda \to \tilde{S}\lambda$,
described by the Feynman diagrams of fig.~3. 
\begin{figure}
\centerline{
\epsfig{figure=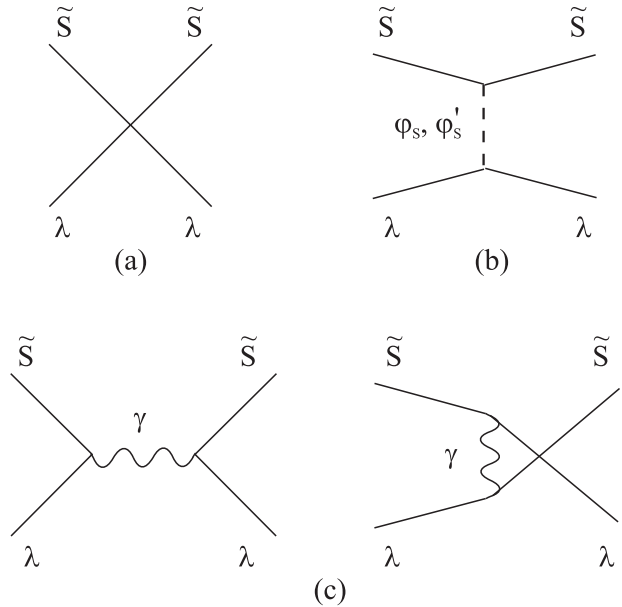,height=8cm,angle=0}
}
\caption{\it Tree--level Feynman diagrams contributing to the
goldstino-gaugino scattering amplitudes: (a) contact interactions;
(b) $(\varphi_S, \varphi_S')$ exchange; (c) photon exchange.}
\label{figtre}
\end{figure}
The Lagrangian terms (A.20) and (A.21) generate the following contact 
interactions (fig.~3a):
\be
{1 \over 8} \left[ - {M^2 \over F^2} +
\left\langle { \fh_{SS} - \fh_S \Gamma^{S}_{SS} \over 
{\rm Re \,} \fh \, \kh_{\ov{S} S} } 
\right\rangle \right] \tilde{S} \tilde{S} 
\lambda \lambda + {\rm h.c.} \, ,
\label{ssll1}
\ee
\be
- {1 \over 2}
{|M|^2 \over |F|^2}
\tilde{S} \lambda
\ov{\tilde{S}} \ov{\lambda}
\, .
\label{ssll2}
\ee
These interactions should be considered together with diagrams 
with virtual ${\varphi_S}, {\varphi_S'}$ exchange (fig.~3b), 
controlled by the cubic couplings (\ref{t2}), (\ref{t3}), and 
diagrams with virtual gauge-boson exchange (fig.~3c), controlled 
by the cubic coupling (\ref{dipole}). We start by discussing the 
low-energy properties. At low energy  ($s \ll m_S^2, m_{S'}^2$), 
the scalars ${\varphi_S}, {\varphi_S'}$ can be easily integrated 
out, and the remaining leading effective interaction is
\be
{1 \over 8} \left[ {4 M^2 \over F^2} -
\left\langle { \fh_{SS} - \fh_S \Gamma^{S}_{SS} \over 
{\rm Re \,} \fh \, \kh_{\ov{S} S} } 
\right\rangle \right] \tilde{S} \tilde{S} 
\lambda \lambda + {\rm h.c. }
\label{ssll3}
\ee
This combines with (\ref{ssll1}) to leave, after a 
major cancellation, the residual term
\be
{3 \over 8} { M^2 \over F^2} \tilde{S} \tilde{S} 
\lambda \lambda + {\rm h.c. }
\label{ssll4}
\ee
On the other hand, the inclusion of the contributions from the 
massless gauge-boson exchange is less simple. In order to
discuss the potential cancellations, we focus on the scattering 
process\footnote{In the annihilation processes, cancellations are 
less effective, since the goldstino energy in the centre-of-mass
frame is bounded from below by the absolute value of the gaugino 
mass, and the low-energy theorems of supersymmetry \cite{let} are
not directly applicable.} ${\tilde S} \lambda \rightarrow {\tilde S} 
\lambda $, for which we can define a Thomson-like low-energy limit, 
where the incident goldstino has energy $E \ll |M|$ in the centre-of-mass 
frame, thus $s \simeq u \simeq |M|^2$ and $t \simeq 0$. Notice 
that, for the different helicity choices, the scattering amplitudes 
receive contributions from gauge-boson exchange only in the $s$ and $u$ 
channels. Therefore, in the kinematical limit considered here, the 
denominators of the propagators are just $|M|^2$, and combine with 
similar factors coming from the attached cubic vertices (\ref{dipole}). 
An explicit evaluation shows that the contributions from the contact terms
(\ref{ssll2}) and (\ref{ssll4}) (the latter corresponds here to 
$t$-channel ${\varphi_S}, {\varphi_S'}$ exchange\footnote{Notice that, 
since $t$ vanishes as $E^2$ in the scattering process considered here, 
the denominators of the ${\varphi_S}$ and ${\varphi_S'}$ propagators
coincide automatically with the corresponding masses, so that it is not 
necessary to assume $s \ll m_S^2, m_{S'}^2$: $E^2 \ll m_S^2, m_{S'}^2$ is
enough.}) are cancelled by $s$- and $u$-channel gauge-boson exchange. As 
a result, the leading amplitudes scale with $E^2 |M|^2/|F|^2$ instead 
of growing linearly with $E$, and the corresponding cross sections scale 
as $\sigma \sim E^4 |M|^2/|F|^4$. We now comment on the high-energy 
r\'egime. Here the contact terms (\ref{ssll1}) and (\ref{ssll2}) compete 
again with contributions from virtual gauge-boson exchange, which are not
suppressed because the propagator can be compensated by the derivative 
cubic vertices: however, no special cancellations occur in general, 
given the different kinematical dependences of the contact interactions
and of the photon-exchange diagrams. For example, in the process
$\tilde{S}_- \lambda_- \to \tilde{S}_- \lambda_- $, which involves
only left-handed goldstinos and gauginos, the contribution to the
amplitude from the contact term (\ref{ssll2}) is $s |M|^2 / 
(2 |F|^2)$, whereas the leading contribution from photon-exchange
diagrams is $(s+2 t) |M|^2 / (2 |F|^2)$. The overall contribution
is $(s + t) |M|^2 / |F|^2 = s \cos^2 (\theta/2) |M|^2 / |F|^2$:
once again, one can derive from unitarity a critical energy $s_c
\sim |F|^2 / |M|^2$, and from the requirements $s_c > m_{\varphi}^2,
|M|^2$ bounds such as $|M|^2, |m_{\varphi} M| < {\cal O} (10) |F|$. 

\subsubsection{$\lambda \lambda \lambda \lambda$}

Before moving to the $A$ sector, we briefly consider also
the scattering process involving four gauginos, i.e. $\lambda 
\lambda \rightarrow \lambda\lambda$, described by the Feynman 
diagrams of fig.~4. 
\begin{figure}
\centerline{
\epsfig{figure=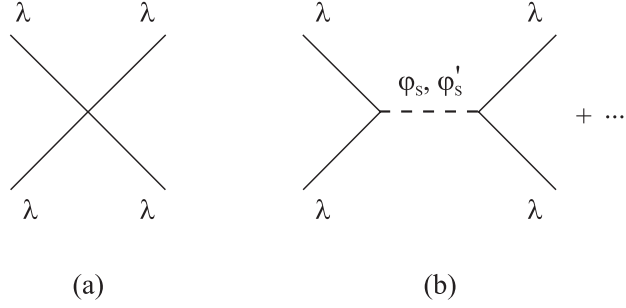,height=4cm,angle=0}
}
\caption{\it Tree--level Feynman diagrams contributing to the
four-gaugino scattering amplitudes: (a) contact interactions;
(b) $(\varphi_S, \varphi_S')$ exchange (the dots stand for the 
$t$- and $u$- channel diagrams.}
\label{figqua}
\end{figure}
The Lagrangian term (A.22) generates the 
contact interaction (fig.~4a)
\be
- {1 \over 4}
{|M|^2 \over |F|^2}
\lambda \lambda \ov{\lambda} \ov{\lambda}
\, ,
\ee
which is again controlled by a model-independent coupling.
At low energy ($s \ll m_S^2, m_{S'}^2$), this interaction should 
be considered together with diagrams with virtual ${\varphi_S}, 
{\varphi_S'}$ exchange (fig.~4b), controlled by the cubic couplings 
(\ref{t2}). However, no special cancellation emerges in this case, 
where no goldstinos are involved. At high energy, the contact 
interaction dominates and produces an amplitude growing as $E^2 |M|^2
/|F|^2$ and a cross section growing as $\sigma \sim s |M|^4/|F|^4$. 
Once again, unitarity is broken for $s\gg |F|^2/|M|^2$ and the 
consistency condition that the critical energy be larger 
than $|M|$ leads to a bound $|M|^2 < {\cal O}(10) |F|$, of the 
same form as the previous bound (\ref{bdip}).

\subsubsection{$\tilde{S} \tilde{S} \tilde{A} \tilde{A}$}

Finally, we discuss processes involving two goldstinos
and two ${\tilde A}$ fermions, described by the Feynman
diagrams of fig.~5. 
\begin{figure}
\centerline{
\epsfig{figure=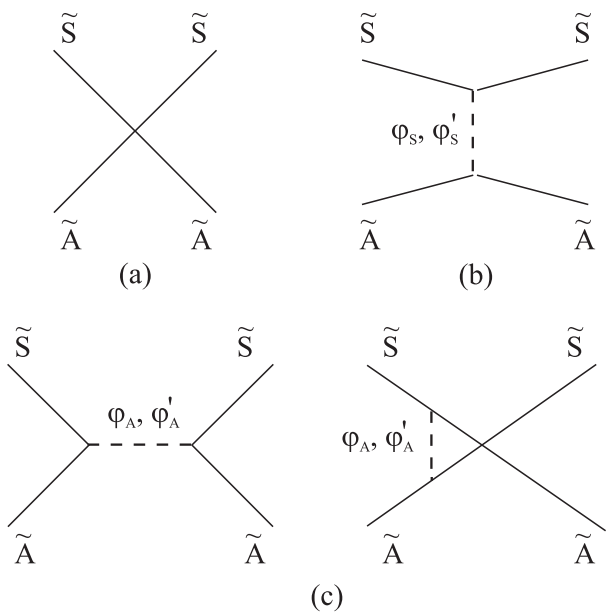,height=8cm,angle=0}
}
\caption{\it Tree--level Feynman diagrams contributing to the
goldstino-$\tilde{A}$ scattering amplitudes: (a) contact interactions;
(b) $(\varphi_S, \varphi_S')$ exchange; (c) $(\varphi_A, \varphi_A')$ 
exchange.}
\label{figcin}
\end{figure}
The Lagrangian term (A.23) generates 
the following contact interactions (fig.~5a):
\be
{1 \over 4} \left\langle { \htil_{SS} - \htil_S \Gamma^S_{SS} 
\over \kt \kh_{\ov{S}S} } \right\rangle \tilde{S} \tilde{S} 
\ov{\tilde{A}} \ov{\tilde{A}} + {\rm h.c.} \, ,
\label{ssaa1}
\ee
\be
- {m_A^2 + m_{A'}^2 - 2 m_{\tilde{A}}^2  \over 2 |F|^2}
\tilde{S} \tilde{A} \ov{\tilde{S}} \ov{\tilde{A}} \, .
\label{ssaa2}
\ee
These interactions should be considered together with diagrams 
with ${\varphi_S},{\varphi'_S}$ exchange (fig.~5b), controlled 
by the cubic couplings (\ref{t3}), (\ref{t6a}) and (\ref{t6b}),
and diagrams with ${\varphi_A},{\varphi'_A}$ exchange (fig.~5c), 
controlled by the cubic coupling (\ref{t7}). We start by discussing 
the low-energy properties. If we integrate out the scalars ${\varphi_S},
{\varphi'_S}$, we obtain the effective interaction 
\be
- {1 \over 4} \left\langle { \htil_{SS} - \htil_S \Gamma^S_{SS} 
\over \kt \kh_{\ov{S}S} } \right\rangle \tilde{S} \tilde{S} 
\ov{\tilde{A}} \ov{\tilde{A}} + {\rm h.c.} \, ,
\label{ssaa3}
\ee
which cancels the above contact term (\ref{ssaa1}). At the same
time, a new term 
\be
{1 \over 8} {m^2_A-m^2_{A'} \over F^2} \tilde{S} \tilde{S} 
{\tilde{A}} {\tilde{A}} + {\rm h.c. }
\label{ssaa4}
\ee
is generated. To continue the discussion of the low-energy limit
and include the ${\varphi_A},{\varphi'_A}$ exchange contributions, 
we consider two different cases. The simplest one corresponds to
assuming $m_{\tilde A} \simeq 0$ and $m^2_{A'} \simeq m^2_A \gg s$. 
Then the term (\ref{ssaa4}) is absent and the remaining term
(\ref{ssaa2}) is exactly cancelled by the leading contribution
from integrating out ${\varphi_A},{\varphi'_A}$. The situation 
is similar to the four-goldstino case: the leading low-energy 
behaviour for the amplitudes has the form $E^4/|F|^2$ instead of 
$E^2 m_A^2/|F|^2$ and the corresponding cross sections scale 
as $\sigma \sim s^3/|F|^2$. Now let us consider a more general
situation, with $m_{\tilde A} \neq 0$. We focus on the scattering process 
${\tilde S} {\tilde A} \rightarrow {\tilde S} {\tilde A}$, since
cancellations are more effective than in the annihilation
processes, thanks to the low-energy theorems of supersymmetry 
\cite{let}. In particular, we consider a Thomson-like 
low-energy limit where the incident goldstino has energy 
$E \ll m_{\tilde A}$ in the centre-of-mass frame\footnote
{In addition, we assume $E \ll |m^2_A-m^2_{\tilde A}|/m_{\tilde A},
|m^2_{A'}-m^2_{\tilde A}|/m_{\tilde A}$.}, so that $s \simeq u 
\simeq m_{\tilde A}^2$ and $t \simeq 0$. Note that, for the different
helicity choices, the scattering amplitudes receive ${\varphi_A}, 
{\varphi'_A}$ exchange contributions only in the $s$ and $u$ channels.
Therefore, in that limit, the propagator denominators contain the same 
scalar-fermion mass splittings appearing in the attached cubic vertices 
(\ref{t7}). The explicit evaluation shows that an amplitude receiving 
a contribution from the contact term (\ref{ssaa2}) is cancelled by 
either $s$- or $u$-channel ${\varphi_A}, {\varphi'_A}$ exchange, whereas 
an amplitude receiving a contribution from the term (\ref{ssaa4}), which 
corresponds here to $t$-channel ${\varphi_S}, {\varphi'_S}$ 
exchange\footnote{Again notice that, since $t$ vanishes as $E^2$ in 
the scattering process considered here, the denominators in the 
${\varphi_S}$ and ${\varphi_S'}$ propagators coincide automatically 
with the corresponding masses, so that it is not necessary to assume 
$s \ll m_S^2, m_{S'}^2$: $E^2 \ll m_S^2, m_{S'}^2$ is enough.}, is 
cancelled by the combined $s$-channel and $u$-channel ${\varphi_A}, 
{\varphi'_A}$ exchange. As a result, the leading amplitudes scale with 
$E^2 m_{\tilde A}^2/|F|^2$ instead of growing linearly with $E$,
and the corresponding cross sections scale as $\sigma \sim E^4 
m^2_{\tilde A}/|F|^4$. Finally, we briefly comment on the high-energy 
r\'egime, considering for simplicity the case in which $m_{\tilde A}\simeq 0$
and the scalars have a common mass $m^2 < s$. Then the situation is similar 
to the case of four-goldstino scattering. The contact interaction 
(\ref{ssaa2}) generates scattering amplitudes that grow as $s m^2/|F|^2$ 
and cross sections that grow as $\sigma \sim s m^4/|F|^4$, so for 
$s \gg |F|^2/m^2$ unitarity breaks down.  The condition that the
critical energy be larger than $m$ leads to a bound of the form 
$m^2 < {\cal O} (10) |F|$, similar to those already discussed in 
the present and previous subsections.

\subsection{Goldstino (light gravitino) phenomenology}

The results of the present section can easily be extended from
the class of models we have used for illustrative purposes to
realistic models, with gauge group $SU(3) \times SU(2) \times
U(1)$ (or larger), and matter content including, besides the
goldstino supermultiplet, also quark, lepton and other chiral
supermultiplets. In particular, we can consider generalizations 
where the $S$ field remains a gauge singlet, while the $A$ field 
is replaced by a set of charged multiplets (for example the MSSM 
fields), coupled to the gauge sector via `usual' gauge 
interactions\footnote{A similar approach was recently followed 
in ref.~\cite{ckn}. However, the emphasis there was on the possibility 
of identifying the scales of flavour and supersymmetry breaking, 
both taken to be of order 10~TeV, preserving at the same time the 
naturalness of the lightest Higgs mass. A more radical generalization, 
where $S$ itself belongs to a charged multiplet, will be discussed 
in the next section.}. As explained in the introduction, our
effective theory description, based on spontaneously broken $N=1$
global supersymmetry, will be adequate as long as the goldstino 
couplings to the supercurrent are larger than the gravitational ones,
that is $\lam_S^4 < \Delta m^2 \mpl^2$, and we consider experiments 
performed at energies $E$ larger than the gravitino mass and
smaller than the unitarity bounds, that is $(\lam_S^4 / \mpl^2) 
\simlt E^2 \simlt (\lam_S^4 / \Delta m^2)$.

To put things in perspective, we recall that, when discussing
the phenomenology of supersymmetric extensions of the Standard
Model, under the assumption of supersymmetry-breaking mass 
splittings of the order of the electroweak scale, it is useful
to distinguish three main possibilities

{\em a) Heavy gravitino.} \hspace{0.15cm}
If the scale of supersymmetry breaking is the geometrical 
mean of the electroweak and Planck scales, the gravitino 
mass is of the order of the electroweak scale, and the 
gravitino couplings to all the other fields have gravitational 
strength. This is the case of the so-called hidden sector
supergravity models. The effective theory near the electroweak 
scale \cite{can} is the MSSM \cite{mssm}, a renormalizable theory with
explicit but soft \cite{gg} breaking of supersymmetry. The additional 
non-renormalizable operators of the underlying supergravity
theory, involving the gravitino and associated with the 
spontaneous breaking of local supersymmetry, are suppressed 
by inverse powers of the Planck mass. Such a property also allows 
a consistent  extrapolation of the MSSM up to a possible scale of 
supersymmetric grand unification, since tree-level unitarity is 
violated only close to the Planck scale. As for the phenomenology 
at colliders, this is the most studied case (see, e.g., 
ref.~\cite{pdb}): the lightest supersymmetric particle is, in 
most of the cases, an admixture of neutral gauginos and higgsinos,
or neutralino, and the footprint of supersymmetric particle 
production is the missing energy associated with the undetected 
neutralino.

{\em b) Light gravitino.} \hspace{0.15cm}
If the scale of supersymmetry breaking is only a few orders
of magnitude above the electroweak scale, say between 10 
and 100 TeV, then the gravitino mass is in the eV--keV range, 
and the goldstino couplings to the MSSM fields have strength
much greater than the gravitational one. This is the case of 
most models of dynamical supersymmetry breaking currently under 
study \cite{recent}. For most practical purposes, we can still take
the MSSM as an effective theory near the electroweak scale, since
the additional operators involving the goldstino and associated 
with the spontaneous breaking of global supersymmetry
are considerably suppressed with respect to the renormalizable MSSM
operators. However, in this case we cannot in general extrapolate 
the MSSM up to a possible scale of supersymmetric grand unification, 
since tree-level unitarity is violated at a much smaller scale.
To make such an extrapolation possible, we have to include the
(model-dependent) degrees of freedom that play a r\^ole in the
restoration of unitarity: in the so-called messenger models 
\cite{recent}, for example, this r\^ole is played by the messenger 
fields. As for the phenomenology at collider energies accessible at present, 
if the lightest supersymmetric particle besides the gravitino is a 
neutralino with non-negligible photino component, 
then the only modification with respect to the previous case 
is due to the effects of the operator (\ref{dipole}), which allows 
for the decay of the lightest neutralino into a photon plus a
goldstino (light gravitino). In this case, the footprints of 
supersymmetric particle production are both missing energy and 
photons, with a photon spectrum dictated by the mass of the 
lightest neutralino. There is also room for a sfermion to be 
the lightest supersymmetric particle besides the gravitino:
in such a case operators of the form (\ref{t7}) control its decays
into the corresponding fermion plus a goldstino (light gravitino),
and we fall back to more conventional missing energy signals,
with the only difference that the undetected particle is now the
gravitino. Recent work on supersymmetric phenomenology at colliders 
in the presence of a light gravitino can be found in 
refs.~\cite{lgpheno}.

{\em c) Superlight gravitino.} \hspace{0.15cm}
If the scale of supersymmetry breaking is very close to the
electroweak scale, then the gravitino is superlight, with a
mass in the $10^{-5}$ eV range, and the goldstino couplings 
to the MSSM fields have a strength comparable to the 
renormalizable MSSM couplings\footnote{There are not many explicit
models in this class, even if this possibility has been mentioned
in the literature \cite{luty}.}. In this case, in order to discuss 
the phenomenology at the electroweak scale we must consider at least 
the full effective theory discussed in the present paper, taking
also into account its limitations when we approach the unitarity
bounds. It is also clear that the energy range of validity of our
effective theory description is very limited, and we should replace it
with a more fundamental theory not much above the electroweak scale.
The interesting point of such a possibility is that it could give
rise, at the present collider energies, to a phenomenology that is
radically different from the previous two cases. Such phenomenology 
should also allow us to establish model-independent lower bounds on 
the scale of supersymmetry breaking, and therefore on the gravitino 
mass, without making reference to the actual masses of the supersymmetric 
particles in the MSSM. For example, we could consider scattering 
processes where the initial state is made of ordinary particles (quarks, 
leptons, gauge bosons), whereas the final state contains both ordinary 
particles and goldstinos (superlight gravitinos), and use the 
phenomenological bounds on the corresponding rates.

In the following, we will discuss some aspects of the superlight
gravitino phenomenology that are strictly related to the results
of the present section.

\subsubsection{Collider signatures of a superlight gravitino}

At $e^+ e^-$ high-energy colliders, such as LEP or the proposed
Next Linear Collider, we could consider processes such as $e^+ e^- 
\to \tilde{S} \tilde{S} \gamma$ or $e^+ e^- \to e^+ e^- \tilde{S} 
\tilde{S}$, and similar processes could be considered at hadron 
colliders, such as the Tevatron or the LHC. Indeed, this was already 
done in the past \cite{goldphen}, but always under special assumptions 
on the supersymmetric particle spectrum, such as light gauginos or 
light spin-0 superpartners of the goldstino: the resulting limits 
on the gravitino mass, of the order of $10^{-6} \ev$, are therefore 
model-dependent. To perform a general and reliable analysis of the 
above processes, including the latest experimental data from LEP and 
the Tevatron collider, would require a considerable amount of work: 
we should generalize the formalism, compute the relevant cross sections 
in the general case, scan the different ranges of supersymmetric particle 
masses, and compare the results with experimental data and background 
calculations. We postpone all this to future work.
We can already anticipate, however, that the weakening of the old 
bounds due to the greater generality will be approximately compensated
by the higher energies accessible at the present machines.

\subsubsection{The processes $\gamma \gamma \longleftrightarrow \tilde{S} 
\tilde{S}$}

The formalism developed in this paper is already sufficient 
to comment on some cosmological and astrophysical 
bounds on the gravitino mass, which were previously studied 
in connection with the annihilation processes $\gamma \gamma
\to \tilde{S} \tilde {S}$ and $\tilde{S} \tilde{S} \to \gamma 
\gamma$. To set the framework, we have computed the general 
expression for the total cross-section of the two processes: 
\be
\label{sigma}
\sigma(\gamma \gamma \to \tilde{S} \tilde{S}) =
\sigma(\tilde{S} \tilde{S} \to \gamma \gamma) =
{|M|^6 \over 32 \pi |F|^4} \left( C_0 + C_1 \log
{|M|^2 \over s + |M|^2} \right) \, ,
\ee
where
\bea
C_0 
& = & 
{m_S^8 \over 4 (s - m_S^2)^2 |M|^4 }
+
{m_{S'}^8 \over 4 (s - m_{S'}^2)^2 |M|^4}
+
{m_S^4 (m_S^2 + 2  |M|^2) \over 2 (s - m_S^2) |M|^4}
+
{m_{S'}^4 (m_{S'}^2 + 2 |M|^2) \over 2 (s - m_{S'}^2) |M|^4}
\nonumber \\ & & \nonumber \\
& + & 
{ m_S^4 + m_{S'}^4 + 4 (m_S^2  + m_{S'}^2) 
|M|^2 +4 |M|^4 \over 4 |M|^4 }
- 
{4 |M|^2 \over s}
+
{|M|^2 \over s + |M|^2}
+
{s \over 6 |M|^2}
\, , 
\label{czero}
\eea
and
\be
\label{cuno}
C_1 
=
{m_S^2 \over s - m_S^2}
+
{m_{S'}^2 \over s - m_{S'}^2}
+
{|M|^2 \over s + 2 |M|^2}
- 
{4 |M|^4 \over s^2}
\, .
\ee
As in the previous literature, we have assumed that the
photino is a mass eigenstate: to be completely general, we 
should include the possibility of gaugino--higgsino mixing.
In the special limits $s \gg m_S^2, m_{S'}^2, |M|^2$ and
$m_S^2,m_{S'}^2 \ll s,|M|^2$ our results are in agreement
with the supergravity results of refs.~\cite{bruno} and
\cite{brdue}, respectively. We now focus on the low-energy
limit, $s \ll |M|^2,m_S^2, m_{S'}^2$, where the expressions
for the scattering amplitudes in the individual helicity 
channels and for the total unpolarized cross-section simplify 
considerably. Indeed, the leading contributions to the
amplitudes and the cross-section become independent of the
the scalar and gaugino masses, as anticipated in section 2.3.2
and summarized by the effective operator of eq.~(\ref{effop}). 
Removing an irrelevant overall phase, and denoting by ${\cal A} 
(\lambda_1, \lambda_2, \sigma_1, \sigma_2)$ the scattering 
amplitude for photon helicities $\lambda_{1,2}$ and goldstino 
helicities $\sigma_{1,2 \;}$, we find
\be
\label{ampli}
{\cal A} (+1,-1;+1/2,-1/2) =
{\cal A} (-1,+1;-1/2,+1/2) =
8 \sin \theta \cos^2 (\theta/2) E^4 / |F|^2 \, ,
\ee
\be
{\cal A} (+1,-1;-1/2,+1/2) =
{\cal A} (-1,+1;+1/2,-1/2) =
- 8 \sin \theta \sin^2 (\theta/2) E^4 / |F|^2 \, ,
\ee
where $E$ and $\theta$ are the energy of each particle
and the scattering angle in the centre-of-mass frame,
respectively. The remaining helicity channels have amplitudes
proportional to higher powers of $E$. The total cross-section 
then becomes:
\be
\label{sigmale}
\sigma(\gamma \gamma \to \tilde{S} \tilde{S}) =
\sigma(\tilde{S} \tilde{S} \to \gamma \gamma) =
{s^3 \over 640 \pi |F|^4} + \ldots \, .
\ee
Here, as already stated, our results differ from those of
ref.~\cite{ghe}. We would like to add that the low-energy 
limit allows also to relax the assumption that the photino be 
a mass eigenstate. In particular, we have checked that the 
results (\ref{effop}) and (\ref{ampli})--(\ref{sigmale}) remain 
valid in the realistic case where the gauge group is the Standard 
Model one and neutralino mixing is taken into account, at least 
if the goldstino belongs to a gauge singlet chiral 
superfield, the gauge kinetic function factorizes
with respect to the gauge group factors, and the 
$D$-terms have vanishing vacuum expectation values.

Very light and stable gravitinos can affect the standard
nucleosynthesis scenario \cite{mmy,ghe, gmr}: if the 
decoupling temperature of goldstinos from the early 
universe thermal bath is sufficiently small, gravitinos 
can contribute to the number $N_{\nu}$ of effective 
neutrino species, which is bounded from above by 
cosmological analyses \cite{nucleo}. The corresponding 
bound on the gravitino mass strongly depends on the
assumed bound on $N_{\nu}$, $N_{\nu}^{max}$, and may
be more stringent than particle physics bounds only
if $N_{\nu}^{max}$ is significantly smaller than 4.
In previous analyses, the dominant reaction affecting
the gravitino thermal equilibrium was considered to be
the annihilation process $\gamma \gamma \leftrightarrow
\tilde{S} \tilde{S}$, on the basis of a cross section 
assumed to grow as $\sigma \sim s^2 |M|^2 / |F|^4$. 
However, if there are no superlight non-standard particles 
(with mass $\simlt 100 {\; \rm MeV}$) besides the gravitino, 
the relevant cross-section is that of eq.~(\ref{sigmale}): 
this means that even the bounds of \cite{ghe,gmr}, which 
in turn were considerably weaker than previous ones 
\cite{mmy}, should be revised. For example,
using $N_{\nu} < 3.6$, ref.~\cite{ghe} found 
$m_{3/2} \simgt 10^{-6} \ev$ (for $|M| \sim 100 \gev$),
whereas on the basis of (\ref{sigmale}) we estimate a
bound about a factor $10^{1-2}$ weaker; to be more precise,
we would need to consider all the reactions that now become
competitive with $\gamma \gamma \leftrightarrow \tilde{S}
\tilde{S}$, but the bounds now become so weak, compared with 
those coming from high-energy accelerators, that the effort is 
probably unjustified.

For very light gravitinos and heavy $(\lambda,S,S')$, 
the annihilation process $\gamma \gamma \to \tilde{S} 
\tilde{S}$ can also be relevant for stellar cooling 
\cite{mmy,ghe,gmr} (other processes that may contribute
to stellar cooling were considered in \cite{fs}). Again, 
using a cross-section going as $\sigma \sim s^2 |M|^2 / 
|F|^4$ and the observed neutrino luminosity from Supernova 
1987a, ref.~\cite{gmr} claimed to exclude the range $2.6 
\times 10^{-8} \ev < m_{3/2} < 2.2 \times 10^{-6} \ev$.
Using eq.~(\ref{sigmale}), we estimate that the
bounds of \cite{gmr} should be reduced by a factor 
of about $10^{1-2}$, losing in part their interest
when compared with possible accelerator bounds.   
Establishing a precise bound, however, would require
the examination of all other processes that may 
become competitive with the above annihilation as a mechanism 
for stellar cooling.

\subsection{Explicit models}

We will now present some explicit and simple models of the 
type discussed above, not only to see the previous general 
considerations at work on some specific example, but also to 
prove the existence of models whose vacuum has the general 
properties discussed so far. For simplicity, in the chiral 
superfield sector we put initially $A$ to zero, and consider 
only~$S$.

\subsubsection{Canonical kinetic terms}

To begin with, we consider the simple case of canonical kinetic
terms, $K = |S|^2$ and $f = 1 / g^2$. The unique superpotential
that leads to a supersymmetry breaking vacuum is then $w =
\Lambda_S^2 \, S$, where $\lam_S$ should be interpreted as some
dynamical scale, induced in the effective theory by non-perturbative
phenomena of the microscopic theory. In such a model, the scalar
potential is just a constant: $V = ||F||^2 \equiv \lam_S^4$. 
Thus $V$ is minimized for arbitrary values of $S$, and all the 
configurations along this complex flat direction break spontaneously 
global supersymmetry, with order parameter $\lam_S$. The uniqueness of 
the supersymmetry-breaking superpotential linear in $S$, in the
case of canonical kinetic terms considered here, is easily
proved. For a generic superpotential $w(S)$, $V = |w'(S)|^2$, and,
from elementary theorems in complex analysis, $V$ can have
a positive relative minimum if and only if $w'(S) \equiv c \ne
0$, so that $w(S) = a S + b$, with $b$ irrelevant for global
supersymmetry.

\subsubsection{Embedding in supergravity}

The previous model can be viewed as an ultra-minimal example
exhibiting the spontaneous breaking of {\em global} supersymmetry.
The physics remains trivial in spite of supersymmetry breaking,
because the canonical choices for $K$ and $f$ do not lead to any 
masses or interactions. However, we will see below that a non-trivial 
physical content can be easily obtained by departing from canonicity. 
Before doing that, we will briefly comment on a {\em locally} 
supersymmetric generalization of the above model, where one of 
the ingredients is also a certain departure from canonicity.
Indeed, the above model can be easily obtained, by taking the 
appropriate low-energy limit, from a corresponding non-trivial 
supergravity model where supersymmetry is spontaneously broken and in 
addition the vacuum energy vanishes. Normalizing the fields conveniently, 
the underlying supergravity model can be defined via the 
$SU(1,1)/U(1)$ K\"ahler potential
\be
K = - 3 \, \mpl^2 \log \left( 1 - 
{1 \over 3} {|S|^2 \over \mpl^2} \right)
\label{ksugs}
\ee
and the superpotential
\be
w =
{ \mpl \lam_S^2 \over \sqrt{3} } \left( 1 +  
{S  \over \sqrt{3} \mpl} \right)^3 \, .
\label{wsugs}
\ee
The scalar potential, which now includes the gravitational contribution, 
is identically vanishing: $V = ||F||^2-||H||^2 \equiv 0$. Comparing 
with the global case, we see that the welcome effect of the gravitational
interactions is the removal of the constant term $\Lambda_S^4$ and
the consequent vanishing of the `true' vacuum energy. Notice that,
if one is interested in the $\mpl \to \infty$ limit only, the
globally supersymmetric model can be directly derived from the
supergravity one by keeping only the leading terms in the $1/\mpl$
expansion of $K$ and $w$:
\be
K = |S|^2 + {\cal O} \left( {S^4\over \mpl^2} \right) \, ,
\ee
\be
w = { \mpl \lam_S^2 \over \sqrt{3} } + \lam_S^2 \, S
+ {\cal O} \left( {\lam_S^2 S^2 \over \mpl} \right) \, ,
\ee
and by discarding the
constant term in the superpotential. We finally remark that, when
the limit $\mpl \to \infty$ with $\lam_S$ fixed is taken, the
gravitino  mass $m_{3/2}^2\equiv\langle e^{K/\mpl^2} |w|^2 /\mpl^4
\rangle$ vanishes as $m_{3/2}\sim \lam_S^2/\mpl$, whereas the
product $m_{3/2} \mpl \sim \lam_S^2$ remains constant and
identifies the (fixed) supersymmetry breaking scale.

\subsubsection{Non-canonical kinetic terms}

We now go back to the globally supersymmetric model considered
above and discuss how more general versions can be obtained
using non-canonical gauge kinetic function and K\"ahler potential. 
Of course, even the simple model considered above can be reformulated 
using non-canonical functions. For example, the analytic field redefinition 
$S = (S')^2/\lam_S$ would transform the superpotential into 
$w = \lam_S (S')^2$ and the K\"ahler potential into 
$K = |S'|^4 / \lam_S^2$, therefore inducing a fake singularity of 
the K\"ahler metric at the origin. However, as guaranteed 
by general theorems, the physical content of the model would 
remain the same. In the search for less trivial versions, one
can consider for instance K\"ahler potentials that  admit a 
power expansion around the canonical form. 

A minimal non-trivial example can be built by keeping the form
$w = \Lambda_S^2 \, S$ for the superpotential and modifying the
K\"ahler potential and gauge kinetic function as follows:
\be
K = |S|^2 - {| S - v |^4 \over 4 \lam^2} \, ,
\label{kfers}
\ee
\be
f = {1 \over g^2} \left( 1 + { S-v \over \lam_f} \right) \, .
\label{fsimple}
\ee
Here $v$, $\lam$ and $\lam_f$ are real and positive mass parameters,
which control the deviations from the canonical forms of $K$ and $f$, 
recovered in the limit $\lam, \lam_f \to \infty$. The choice 
(\ref{kfers}) leads to a K\"ahler metric $K_{\ov{S} S} = 1 - |S-v|^2 
/ \lam^2$, so that the allowed field configurations, corresponding to 
a positive-definite K\"ahler metric, are those satisfying the inequality
$|S-v| < \lam$. Thus the scale $\lam$, which controls the 
deviations from a canonical K\"ahler potential, acts also as 
an effective cutoff in field space. The scalar potential is now
\be
\displaystyle{||F||^2 = \displaystyle{{\lam_S^4 \over 1 
- \displaystyle{{|S-v|^2 \over \lam^2}} }}} \, .
\ee
It is manifestly positive-definite in the domain of validity of the
model and has an absolute minimum for $\langle S \rangle =v$, 
where $\langle ||F||^2 \rangle = \lam_S^4$. Supersymmetry is 
spontaneously broken with order parameter $\lam_S$ as before 
and the gauge and K\"ahler metrics are canonical only at the 
minimum. Since the $S$ direction is no longer flat, the scalars 
$\varphi_S$ and $\varphi_S'$ now have non-vanishing masses,
\be
m_S^2 = m_{S'}^2 = {\lam_S^4 \over \lam^2} \, ,
\label{phim}
\ee
controlled by the supersymmetry-breaking scale and the curvature
of the K\"ahler manifold at the minimum of the potential.
Similarly, a non-vanishing gaugino mass is generated,
\be
\label{gausimple}
M = {\lam_S^2 \over 2 \lam_f} \, ,
\ee
controlled by the supersymmetry-breaking scale and $\lam_f$.

Besides generating non-vanishing masses, the non-canonical
$K$ and $f$ also generate non-trivial interactions, which
are special cases of the ones discussed in the previous 
subsections. In particular, the general 
bounds (\ref{bsgf}), (\ref{bsss}) and (\ref{bdip}) relating 
masses and $|F|$ can be converted into the following bounds 
relating $\lam_S$, $\lam$ and $\lam_f$:
\be
\lam_S^2 < 2\sqrt{32 \pi} \lam \lam_f \, ,
\;\;\;\;\;
\lam_S^2 < \sqrt{32 \pi} \lam^2 \, , 
\;\;\;\;\;
\lam_S^2 < 4 \sqrt{16 \pi} \lam_f^2 \, .
\label{lll}
\ee
We recall that such relations come from the requirement
that certain effective cubic couplings be not too large.
Similarly, one can specialize to the present case the 
unitarity bounds derived from the study of four-particle 
processes. For instance, the critical energy (\ref{scri}),
which signals the unitarity breakdown in the four-goldstino
amplitudes, now becomes directly related to the scale $\lam$:
\be
\sqrt{s}_c=\sqrt{24\pi} \lam \, .
\ee
As a consequence, the requirement $s_c>m_S^2+m_{S'}^2$, leading 
to (\ref{bscri}), now gives 
\be
\label{lsl}
\lam_S^2 < \sqrt{12 \pi} \lam^2 \, ,
\ee
which is consistent with the second inequality in (\ref{lll}).

As a final remark, we note that the $A$ field could be easily 
added to the above toy model. In particular, a mass splitting
in the $A$ sector can be generated by choosing a non-canonical
metric ${\tilde K}(S,{\ov S})$, in the notation of (\ref{kappc}).
For example, the simple choice
\be
{\tilde K} = 1 - {| S - v |^2 \over  {\tilde \lam}^2 } \, 
\ee
leads to a metric that is canonical only at the minimum and 
that induces, according to eq. (\ref{softm}), scalar soft masses 
\be
{\tilde m}^2_A = {\lsu^4 \over {\tilde \lam}^2} \, .
\ee
In particular, the scales $\lam_S$, $\lam$, $\lam_f$ and 
$\tilde{\lam}$ may coincide with a single dynamically generated
scale of an underlying microscopic theory and still remain 
compatible with the bounds (\ref{lll}) and (\ref{lsl}).
\nsect{The case of a non-singlet goldstino}

In this section, we will examine some special features of 
models where the spontaneous breaking of $N=1$ global
supersymmetry is strictly related with the spontaneous 
breaking of the gauge symmetry, and the goldstino is charged
under the broken gauge group. Our aim is to show that, while
most of the general results of the previous section can be
easily adapted to the new framework, the constraints coming 
from gauge invariance lead to an increased predictive power
and to new effects. 
In particular, we will show that the singularity structure
of the defining functions of the model, needed to spontaneously 
break both supersymmetry and gauge symmetry, 
leads to a special class of interactions, controlled by
the scales of gauge and supersymmetry breaking, and that the
latter are subject to peculiar constraints from tree-level 
unitarity.

For our purpose, it will be sufficient to consider the class
of $N=1$ globally supersymmetric models with gauge group 
$G=U(1)_Y$ and two chiral superfields $(H_1,H_2)$ of opposite 
non-vanishing charges, conventionally normalized to $Y(H_1)=-1/2$ 
and $Y(H_2)=+1/2$. We assume that no Fayet--Iliopoulos term is
present: thanks to the non-anomalous field content of the theory,
this situation is stable against perturbative quantum corrections. 
We start with two very simple observations, which have important 
implications. First, if supersymmetry is spontaneously broken, 
then the gauge symmetry must be broken at the same time, since 
these models do not contain neutral superfields. 
Second, if the superpotential $w$, the gauge kinetic function 
$f$ and the K\"ahler metric $K_{\ov{\j}i}$ have no singularities 
(including zeros) at the origin, so that they can be Taylor-expanded 
around $H_1=H_2=0$, then supersymmetry is unbroken. In fact,
gauge invariance forces the first terms in the Taylor expansions of
$w$ and $K$ to be at least quadratic in the fields; thus both $F_i
\equiv w_i$ and $D_Y \equiv Y_{(i)} K_i \varphi^i$ vanish at the 
origin, and so does the potential $V = ||F||^2 + ||D||^2$.
This important result is indeed valid for any $N=1$ globally 
supersymmetric model that does not contain gauge-singlet chiral 
superfields nor a Fayet--Iliopoulos term, irrespective of the gauge 
group and of the representation content: a necessary condition for 
supersymmetry breaking is the singular behaviour at the origin 
of one of the defining functions of the model. In the following, 
for definiteness, we will consider models where the singularity 
at the origin comes from the superpotential, whereas the K\"ahler 
potential and the gauge kinetic function admit a power expansion 
around their canonical form. 

We would like to study the peculiar features of our class of models
without excessive complications, limiting ourselves to the
case of pure $F$-term supersymmetry breaking.
Notice that the basic functions $w$ and $f$ are necessarily
symmetric under $H_1 \leftrightarrow H_2$, since they must
be functions of the only available analytic invariant $H_1 H_2$.
For simplicity, we will assume that $K$ is symmetric as well.
For the moment, we do not discuss the 
minimization of the scalar potential: we just assume the existence of 
a minimum that breaks both gauge symmetry and supersymmetry,
with $\langle || F ||^2 \rangle \ne 0$, $\langle || D ||^2 \rangle = 0$.
Notice that in general the minimum of the potential can be parametrized by
the three real gauge-invariant variables $v_1$, $v_2$ and $\theta$, defined by
$\langle |H_1|^2 \rangle \equiv v_1^2$, 
$\langle |H_2|^2 \rangle \equiv v_2^2$ and 
$\langle H_1 H_2 \rangle \equiv v_1 v_2  e^{2 \i \theta}$.
However, our assumptions of symmetric $K$ and no $D$-term breaking imply 
$v_1=v_2 \equiv v/\sqrt{2}$, and we can always gauge-rotate the fields so 
that $\langle H_1 \rangle = \langle H_2 \rangle = v e^{\i \theta}/\sqrt{2}$.
Moreover, in order to make contact with the formalism of the previous section,
it is convenient to introduce the symmetric and antisymmetric combinations 
of fields:
\be
S \equiv {H_1+H_2 \over \sqrt{2}} \, ,
\;\;\;\;\;
A \equiv {H_1-H_2 \over \sqrt{2}} \, .
\label{sas}
\ee
The fields defined in (\ref{sas}) are special cases of the $S$ and $A$ 
fields of section~2. In particular, they will satisfy the basic conditions 
(\ref{cond}), and $S$ will have a non-vanishing vev, $\langle S \rangle 
= v e^{\i \theta}$.

To discuss the mass spectrum around the chosen vacuum, 
it is sufficient to expand the defining functions of 
the model up to quadratic fluctuations in the fields with 
vanishing vevs, as in eqs.~(\ref{wappc})--(\ref{fappc}).
However, now the functions of $S$ appearing as coefficients in the 
$A$-field expansion are not independent,
since $S$ and $A$ originate from the same charged multiplets
and  are linked by gauge invariance\footnote{Of course,
this comment does not apply to additional $A$-type 
fields that can be added to the model, i.e. `matter' fields 
not belonging to the same symmetry multiplet containing $S$.}.
In particular, taking into account that $H_1$ and $H_2$ enter the 
defining functions of the model only via the gauge-invariant bilinears 
$|H_1|^2+|H_2|^2 = |S|^2+|A|^2$ and $H_1 H_2 = (S^2 - A^2)/2$,
the following field identities can be derived:
\be
\label{iduno}
\wh_S(S) \equiv - S \, \wt (S) \, ,
\ee
\be
\label{iddue}
\kh_S (S,{\ov S}) \equiv \ov{S} \, \kt ( S, \ov{S} ) -
S \, \htil (S, \ov{S}) \, .
\ee
We can then discuss the structure of the mass spectrum
by specializing the general formulae of section 2.1, taking
into account the useful identities (\ref{iduno}) and (\ref{iddue}),
and by including the effects of the standard gauge interactions. 
The mass of the gauge boson is given by:
\be
\label{mgbgen}
m_V^2 = {g^2 v^2 \over 2} \, \langle \kt \rangle \, .
\ee
In the fermionic sector, we can identify on general grounds
the massless goldstino with the symmetric higgsino combination,
$\tilde{S}$. More specifically, the mass matrix has the form:
\be
\label{mfer}
i \lambda,  \tilde{A}, \tilde{S}:
\;\;\;\;\;
{\cal M}_{1/2} = \left(
\begin{array}{ccc}
M & m_V & 0 \\
m_V & \hat{\mu} & 0 \\
0 & 0 & 0
\end{array}
\right) \, ,
\ee
where, using (\ref{iduno}), (\ref{iddue}) and (\ref{minis})
in the general formula (\ref{mferm}),  
\be
\hat{\mu} = - \left\langle {\ov{S} \over S} \ov{F}^{\ov{S}}
\left[ \log \left( \ov{S} \, \kt \right)
\right]_{\ov{S}} \right\rangle \, .
\ee 
The spin-0 $S$ sector contains two physical scalars. In particular, 
if $\langle V_{SS} \rangle = \langle V_{{\ov S}{\ov S}} \rangle$, 
the mass eigenstates are just the real and imaginary parts of the 
shifted $S$ field, $\varphi_S$ and $\varphi_S'$ (\ref{reims}). 
The corresponding mass eigenvalues have the general expressions 
(\ref{smasses}), which can be further specialized if 
additional information on $\wh(S)$ and $\kh(S,\ov{S})$ is given.
The spin-0 $A$ sector contains a physical scalar and the 
unphysical Goldstone boson absorbed by the massive gauge
boson. The fact that the determinant of the $F$-term mass matrix 
vanishes is a consequence of the equality
$m^2_{AA}=m^2_{A {\ov A}}\langle {\ov S}/S \rangle$, which
in turn follows from the use of (\ref{iduno}), (\ref{iddue}) 
and (\ref{minis}) in the general formulae (\ref{mdiag}), (\ref{softm})
and (\ref{moffd}). The physical scalar particle has mass
\be
\label{scalmas}
m_A^2 = m_V^2 + \hat{m}^2 \, ,
\ee
where
\be
\hat{m}^2 = 2\left( |{\hat \mu}|^2 +  \left\langle 
{\ov F}^{\ov S} (- \log\kt)_{\ov{S} S} F^S \right\rangle \right)
\, .
\ee
In particular, if $\langle S \rangle=\langle \ov{S}\rangle$, the 
physical scalar and the Goldstone boson are just the real and imaginary 
parts of $A$, $\varphi_A$ and $\varphi'_A$ (\ref{reima}). 

We should now discuss the interactions of the chosen class
of models. Besides those already studied in section~2, we
should include ordinary gauge interactions and work
out explicitly the interactions associated with the singular 
behaviour of the superpotential at the origin. 
Instead of attempting a general treatment, however, we choose 
to continue the discussion by switching directly to specific examples, 
stressing the properties that have more general validity.
We make this choice both for simplicity and to provide 
`existence proofs' of the non-trivial situation where
supersymmetry and gauge symmetry are broken
simultaneously. The explicit models that we will present
are non-trivial extensions of the corresponding ones of 
section 2.5.

\subsection{Explicit models: canonical kinetic terms}

As in 2.5.1, we begin by considering the simple case of a canonical
gauge kinetic function,
\be
\label{fcan}
f = {1 \over g^2} \, ,
\ee
and a canonical K\"ahler potential,
\be
\label{kcan}
K = |H_1|^2 + |H_2|^2 = |S|^2+|A|^2 \, .
\ee
Then the {\em unique} superpotential that leads to a supersymmetry-breaking
vacuum is
\be
\label{toyw}
w = \lsu^2 \sqrt{2 H_1 H_2} = \lsu^2 \sqrt{S^2 - A^2} \, ,
\ee
where the numerical factor $\sqrt{2}$ is just for convenience
and an irrelevant additive constant has been omitted.
In the above expression, $\lam_S$ should be interpreted
as some dynamical scale, induced in the effective theory by
non-perturbative phenomena of the microscopic theory. The
presence of non-renormalizable interactions in the superpotential
and the lack of analyticity at the origin (conical singularity of
deficit angle $\pi$) should not come as a surprise, given the
existing literature \cite{dsb,seiberg} on non-perturbative 
phenomena in supersymmetric theories.

In the model defined by eqs.~(\ref{fcan})--(\ref{toyw}), the
$F$-term and $D$-term contributions to the scalar potential are
easily computed (see the appendix for the general formulae):
\be
\label{fsusy}
||F||^2 = \lsu^4 { |H_1|^2+|H_2|^2 \over 2 |H_1| |H_2| }
= \lsu^4 + \lsu^4 { (|H_2| - |H_1|)^2 \over 2 |H_1| |H_2|} \, ,
\ee
\be
||D||^2 =
{g^2 \over 8} \left( |H_2|^2 - |H_1|^2 \right)^2 \, .
\ee
It is immediate to see that $V= ||F||^2 + ||D||^2$ is positive 
definite, and is minimized for arbitrary values of $v_1
= v_2 \equiv v/\sqrt{2}$ and $\theta$.  All the configurations
along these two flat directions spontaneously break global
supersymmetry, with order parameter $\langle ||F||^2 \rangle
= \lam_S^4 $, and the $U(1)$ gauge symmetry. Conversely,
it is easy to demonstrate the uniqueness of the superpotential
(\ref{toyw}), given the canonical terms (\ref{fcan})
and (\ref{kcan}). Since the only independent
analytic invariant is $z \equiv H_1 H_2$, the generic
gauge-invariant superpotential can be written as $w(z)$,
as already pointed out. Therefore the scalar potential reads
\be
\label{vtoy}
V = 2 \left| w'(z) \sqrt{z} \right|^2 +
\left( |H_2| - |H_1| \right)^2 \left[
|w'(z)|^2 + {g^2 \over 8} ( |H_2| + |H_1| )^2
\right] \, .
\ee
The two terms in (\ref{vtoy}) are separately positive
semi-definite, and the second vanishes for $|H_1|=|H_2|$.
This leaves the freedom to minimize the first term with respect
to the complex variable $z$. From elementary theorems in
complex analysis, $| w'(z) \sqrt{z}|^2$ can have a positive
relative minimum if and only if $w'(z) \sqrt{z} \equiv c \ne
0$, so that $w(z) = a \sqrt{z} + b$, with $b$ irrelevant for
global supersymmetry.

\subsubsection{Masses and interactions}

The computation of the masses and interactions 
around the generic vacuum is straightforward, and can be easily
obtained by specializing the Lagrangian given in the appendix.
In particular, the results for the spectrum agree with the
general formulae given above. Since the mass eigenvalues
turn out to be independent of $\theta$, for simplicity we will 
expand the Lagrangian around a minimum with $\theta=0$. Therefore the 
chosen vacuum has $\langle H_1 \rangle = \langle H_2 \rangle = 
v/\sqrt{2}$, i.e. $\langle S \rangle=v$ and $\langle A \rangle = 0$.

The crucial feature of the model is the presence,
besides the renormalizable gauge interactions, of a
tower of non-renormalizable interactions, generated by
the expansion of the superpotential (\ref{toyw}) around
the vacuum. The relevant terms of the Lagrangian are
\bea
||F||^2
& = &
\Lambda_S^4 {|S|^2 + |A|^2 \over |S^2 - A^2|}
\nonumber \\ & & \nonumber \\
& = &
\lam_S^4 + {\lam_S^4 \over v^2} \varphi_A^2
+ {\cal O} \left( {\lam_S^4  \over v^3} \varphi^3 \right)
+ {\cal O} \left( {\lam_S^4 \over v^4} \varphi^4 \right)
+ {\cal O} \left( {\lam_S^4 \over v^5} \varphi^5 \right)
+ \ldots \, ,
\label{fterm}
\eea
where we have used the generic symbol $\varphi$ for any of the
fields $(\varphi_A,\varphi_S,\varphi_A',\varphi_S')$, and 
\bea
{\cal L}_{YUK}
& = &
{\lam_S^2 \over 2}
{
(A \tilde{S} - S \tilde{A})^2
\over
(S^2 - A^2)^{3/2}}
+ {\rm h.c.}
\nonumber \\ & & \nonumber \\
& = &
\left[
{1 \over 2} {\lam_S^2 \over v} \tilde{A} \tilde{A}
-  {\lam_S^2 \over v^2}
\left( \phi_A \tilde{S} \tilde{A} +
{1 \over 2} \phi_S \tilde{A} \tilde{A}  \right) \right.
\nonumber \\ & & \nonumber \\
& & \left. + {\lam_S^2 \over v^3}
\left( {1 \over 2} \phi_A^2 \tilde{S} \tilde{S} + 
2 \phi_S \phi_A \tilde{S} \tilde{A} +
{1 \over 2} (\phi_S^2 + 3 \phi_A^2) \tilde{A} \tilde{A}  \right)
+ \ldots \right] + {\rm h.c.} \, ,
\label{yukawa}
\eea
where $\phi_S \equiv S-v \equiv (\varphi_S + i \varphi'_S)/\sqrt{2}$
and $\phi_A \equiv A \equiv (\varphi_A + i \varphi'_A)/\sqrt{2}$.
Among the renormalizable gauge interactions, we write down
those involving fermions:
\be
\label{gaugeint}
{\cal L}' = 
{g \over 2} A_{\mu} {\ov {\tilde A}} {\ov \sigma}^{\mu} {\tilde S}
-{i g \over \sqrt{2}} {\ov \phi}_A {\tilde S} \lambda
-{i g \over \sqrt{2}} (v+{\ov \phi}_S) {\tilde A} \lambda + {\rm h.c.} 
\ee

We begin by summarizing the mass spectrum, part of which is 
already visible in the above expansions. Remembering that, in the 
notation of eq.~(\ref{kappc}), $\tilde{K} \equiv 1$, we find from
(\ref{mgbgen}) that the $U(1)$ gauge boson has a mass:
\be
m_V = {g v \over \sqrt{2}} \, .
\ee
In the spin-0 $S$ sector, we find that $m_S = m_{S'} =0$.
In the spin-0 $A$ sector, eq.~(\ref{scalmas}) holds for the
$\varphi_A$ mass, with
\be
\hat{m}^2 = {2 \Lambda_S^4 \over v^2} \, ,
\ee
whereas $\varphi'_A$ is the unphysical Goldstone boson.
In the fermionic sector, the mass matrix has the form
(\ref{mfer}), with $M=0$ and
\be
\hat{\mu} = - {\Lambda_S^2 \over v} \, .
\ee

The above mass spectrum exhibits a number of interesting
features. Observe first that, since $\langle
|| D ||^2 \rangle = 0$, the mass terms proportional to
$g v$ are distributed in a supersymmetric way, in the
sense that the contributions to $\str {\cal M}^{2n}$
proportional to $(gv)^{2n}$ are identically
vanishing for any positive integer $n$. Indeed, the
supersymmetry-breaking mass splittings are all in
the $(A,\tilde{A})$ sector, and are controlled
by the ratio $\lam_S^4/v^2$; there are no mass splittings
in the $(S,\tilde{S})$ sector. As already mentioned in
section~2.1, this is a consequence
of a general result of global supersymmetry \cite{fgp}:
for canonical kinetic terms and in the absence of anomalous
$U(1)$ gauge factors, $\str {\cal M}^2 \equiv 0$ in each
separate sector of the mass spectrum. In the $(A,\tilde{A})$
sector, where the fermion is massive, the two scalar degrees
of freedom can have a mass splitting. In the $(S,\tilde{S})$
sector, which contains the massless goldstino, no mass splitting
is allowed. Having in mind possible extensions to realistic models,
the existence of massless scalars poses a potential problem.
However, this can be solved by the introduction of non-canonical 
kinetic terms, as we have seen in section~2. Similar
considerations apply to the diagonal gaugino mass term,
which may be needed in realistic extensions of our toy model:
its vanishing is due to the choice of a canonical gauge kinetic
function $f$, but we know that a non-vanishing gaugino
mass term can be generated by a non-canonical $f$.
As a final comment on mass relations, notice that the non-renormalizable 
structure of the superpotential leads to violate a mass sum rule of the 
MSSM type: $m_V^2 + m_{S'}^2 \ne m_A^2 + m_S^2$.

By looking at eqs.~(\ref{fterm}) and (\ref{yukawa}), we can see
that the ratio $\lam_S/v$ simultaneously  controls all the operators
of different dimensionality generated by the superpotential
(\ref{toyw}): the non-vanishing fermion and scalar masses are of
order $\lam_S^2/v$;
the $d=3$ cubic scalar operators have mass parameters of order
$\lam_S^4/v^3$; the $d=4$ quartic scalar couplings and Yukawa
couplings have dimensionless coefficients of order $\lam_S^4/v^4$
and $\lam_S^2/v^2$, respectively; the $d=5$
couplings of the type $\varphi^5$ or $\tilde{H} \tilde{H} \varphi^2$
have dimensionful coefficients of order $\lam_S^4/v^5$ and $\lam_S^2
/ v^3$, respectively, and so on. We then see that three qualitatively
different situations are possible. If $\lam_S \ll v$, then the
supersymmetry-breaking mass splittings are much smaller than the
scale of gauge symmetry breaking, and the interactions induced by
the superpotential are strongly suppressed by positive powers of
$\lam_S/v$. An obvious physical application would be to identify
the broken gauge group with some grand-unified group, as already
attempted elsewhere \cite{kq,bfz} in the supergravity context: the
scale of supersymmetry breaking would then be $\lsu \sim G_F^{-1/4}
M_{\rm GUT}^{1/2} < G_F^{-1/4} \mpl^{1/2}$. If $\lam_S \sim v$,
then the scales of gauge and supersymmetry breaking are the same,
and all the operators in (\ref{fterm}) and (\ref{yukawa}) have
coefficients controlled by the appropriate power of this common
scale: this will be the case, in a realistic framework, if the 
broken gauge group is identified with the Standard Model one. 
As we will see in
a moment, tree-level unitarity strongly restricts the energy range
in which the model is valid. Finally, the case $\lam_S \gg v$
corresponds to a strong coupling r\'egime, where the scale of
supersymmetry breaking is much larger than the scale of gauge
symmetry breaking, and, for example, the dimensionless couplings
of some $d=4$ operators become much larger than~1. To deal
with this case consistently, we should remove the heavy degrees of
freedom and consider a non-linear realization of supersymmetry in
terms of the light degrees of freedom only.

The above discussion shows that, in spite of the canonical $K$ and $f$,
the model under study can enter a strong coupling r\'egime or
violate unitarity, because $w$ generates couplings scaling with
inverse powers of $v$. We can quantify such effects proceeding
as in section~2. For simplicity, we will focus on those couplings
and neglect the renormalizable gauge interactions, i.e. we will 
assume $g \ll \lam_S^2/v^2$. Notice that, since $m_V \ll |{\hat \mu}|$
in this limit, the fermion ${\tilde A}$ is a mass eigenstate with 
mass $|{\hat \mu}|$, whereas the scalar $\varphi_A$ has mass $m_A \simeq 
\sqrt{2} |{\hat \mu}|$. First, we can use two-body decays to 
constrain the parameter space where coupling constants are in the 
perturbative r\'egime and a particle interpretation is allowed.
For instance, we can consider the decay $\varphi_A \rightarrow 
\tilde{S} \tilde{A}$, controlled by one of the cubic Yukawa couplings 
of eq.~(\ref{yukawa}). The calculation goes exactly as in
section~2.2.6, and from the bound of eq.~(\ref{sabound}) we obtain:
\be
\lam_S < 2 \sqrt[4]{4 \pi} v \simeq 4 \, v \, .
\label{lsv} 
\ee
In terms of $m_A$, this means $m_A \simlt 5 \lam_S$, or $m_A \simlt 20 v$. 
Second, we can use four-particle amplitudes to identify the energy 
scale characterizing unitarity violations, which is a cut-off scale
for the effective theory. For instance, we can consider processes
involving two $\varphi_A$ scalars and two goldstinos, i.e.
${\tilde S} \varphi_A \to {\tilde S}  \varphi_A$,
${\tilde S} {\tilde S} \to \varphi_A \varphi_A$ and 
$\varphi_A \varphi_A \to {\tilde S} {\tilde S}$.
In the high-energy limit, the leading contributions to
the scattering amplitudes originate from the corresponding 
contact term of eq.~(\ref{yukawa}):
\be
{\lam_S^2 \over 4 v^3} \varphi_A^2 \tilde{S} \tilde{S}
+ {\rm h.c.}
\label{contca}
\ee
This term generates amplitudes that grow as ${\sqrt s} 
\lsu^2/v^3$, and unitarity breaks down beyond a critical energy
${\sqrt s_c}  = {\cal O} (10) v^3/\lsu^2$. Notice that,
for fixed $v$, $s_c$ decreases for increasing $\Lambda_S$. If we
impose the consistency requirement $s_c > m_A^2$, we obtain
bounds of the form $\lsu \simlt ({\rm few}) v$, as in (\ref{lsv}).

\subsubsection{Embedding in supergravity}

The toy model of eqs.~(\ref{fcan})--(\ref{toyw}) should be
viewed as a sort of minimal example exhibiting the simultaneous
and spontaneous breaking of {\em global} supersymmetry and of
gauge symmetry. In the context of {\em local} supersymmetry,
i.e. supergravity, such an approach was previously studied in
refs.~\cite{kq,bz,bfz}. We recall that, in the presence of gravitational
interactions, the additional property of vanishing vacuum energy 
can be fulfilled at the same time. Indeed, it is instructive to see
how our toy model can be obtained, by taking the appropriate
low-energy limit, from one of the supergravity models with
spontaneously broken supersymmetry and vanishing vacuum energy
considered in ref.~\cite{bfz}. Normalizing conveniently the fields,
the underlying supergravity model can be defined via the 
$(SU(1,1)/U(1))^2=SO(2,4)/SO(2)\times SO(4)$ K\"ahler
potential
\be
K = - {3 \over 2} \mpl^2 \log \left[ \left( 1 -
{2 \over 3} {|H_1|^2 \over \mpl^2} \right) \left(
1 - {2 \over 3} {|H_2|^2 \over \mpl^2} \right) \right]
\label{ksugra}
\ee
and the superpotential
\be
w = { \mpl \lam_S^2 \over \sqrt{3}}
\left( 1 +  \sqrt{ 2 H_1 H_2 \over 3 \mpl^2} \right)^3 \, .
\label{wsugra}
\ee
Since the $D$-term contribution to the scalar potential has the same
form in local and global supersymmetry (see the appendix for the general
formulae), we can concentrate on
\bea
||F||^2 - ||H||^2
& = & \Lambda_S^4 
\frac { \left( 1 +  { 2 |H_1 H_2| \over 3 \mpl^2} \right) \,
\left| 1 +  \sqrt{ 2 H_1 H_2 \over 3 \mpl^2} \right|^4 }
{ \left( 1 - {2 \over 3} {|H_1|^2 \over \mpl^2} \right)^{3/2}
  \left( 1 - {2 \over 3} {|H_2|^2 \over \mpl^2} \right)^{3/2} }
{ ( |H_2| - |H_1|)^2 \over 2 |H_1| |H_2|}
\label{vfull}
\\
& = & \Lambda_S^4 
{ ( |H_2| - |H_1|)^2 \over 2 |H_1| |H_2|}
\left[ 1 + {\cal O} \left( {H^2 \over \mpl^2} \right)
\right] \, .
\label{vappr}
\eea
The first line (\ref{vfull}) shows that $||F||^2-||H||^2$ is positive
semi-definite and vanishes identically along the supersymmetry- and
gauge-symmetry-breaking minima $\langle|H_1|\rangle=\langle|H_2|
\rangle$ (which make $||D||^2$ vanish as well). Of course, this
property is inherited by the leading term (\ref{vappr}) in the
$1/\mpl$ expansion of $||F||^2-||H||^2$. Notice that (\ref{vappr})
coincides with the field-dependent part of (\ref{fsusy}).
The only (welcome) effect of the gravitational interactions,
in the $\mpl \to \infty$ limit, is the removal of the constant
term $\Lambda_S^4$ and the consequent vanishing of the `true'
vacuum energy. Notice also that, if one is interested in the $\mpl
\to \infty$ limit only, the globally supersymmetric model can be
directly derived from the supergravity one by keeping only the
leading terms in the $1/\mpl$ expansion of $K$ and $w$:
\be
K = |H_1|^2 + |H_2|^2 + {\cal O} \left( {H^4\over \mpl^2}
\right) \, ,
\ee
\be
w = { \mpl \lam_S^2 \over \sqrt{3} } + \lam_S^2 \sqrt{2 H_1 H_2}
+ {\cal O} \left( {\lam_S^2 H^2\over \mpl} \right) \, ,
\ee
and by discarding the
constant term in the superpotential. We finally remark that, when
the limit $\mpl \to \infty$ with $\lam_S$ fixed is taken, the
gravitino  mass $m_{3/2}^2\equiv\langle e^{K/\mpl^2} |w|^2 /\mpl^4
\rangle$ vanishes as $m_{3/2}\sim \lam_S^2/\mpl$, whereas the
product $m_{3/2} \mpl \sim \lam_S^2$ remains constant and
identifies the (fixed) supersymmetry-breaking scale.

\subsection{Explicit models: non-canonical kinetic terms}

The globally supersymmetric model described above is clearly 
non-trivial, in spite of the canonical K\"ahler potential 
and gauge kinetic function, and has allowed us to illustrate 
a number of important issues.
We would like to comment here on more general versions
of the model, characterized by non-canonical $K$ and $f$.
Before moving to models with a different physical content,
it may be useful to recall that models equivalent to the
previous one can be easily obtained via analytic field 
redefinitions, which could in particular shift the singularity 
at the origin from the superpotential to the
K\"ahler metric. For example, the redefinition  $H_1 = (H_1')^2/\lam_S$, 
$H_2= (H_2')^2/\lam_S$ would transform the K\"ahler potential into 
$K = ( |H_1'|^4 + |H_2'|^4 )/\lam_S^2$ and the superpotential into 
$w = \sqrt{2} \lam_S H_1' H_2'$, whereas the redefinition
$\phi=\sqrt{H_1 H_2}$, $\phi'=\lam_S \sqrt{H_1/H_2}$ would transform
the K\"ahler potential into
$K = |\phi|^2 (|\phi'|^2/\lam_S^2 + \lam_S^2/|\phi'|^2)$ 
and the superpotential into $w= \sqrt{2} \lam_S^2 \phi$.

We now show a physically different modification of the toy model,
which allows us to lift the flat $S$-direction, i.e. to
fully determine the vacuum and to make the $S$-scalars massive.
In particular, we will keep the form (\ref{toyw}) of the 
superpotential and consider the following modification of the 
K\"ahler potential, which generalizes the modification
(\ref{kfers}) of section 2.5.2: 
\be
K = |H_1|^2 + |H_2|^2 -
{ |\sqrt{2 H_1 H_2} - v|^4 \over  4 \lam^2} \, ,
\label{kfer}
\ee
where $v$ and $\lam$ are real and positive mass parameters.
The condition that the corresponding K\"ahler metric 
\be
K_{\ov{\j} i} =
\left(
\begin{array}{cc}
1 - \left| {H_2 \over H_1} \right|
{ |\sqrt{2 H_1 H_2} - v|^2 \over  2 \lam^2} 
&
-  \sqrt{ H_1 \ov{H}_2 \over H_2 \ov{H_1} }
{ |\sqrt{2 H_1 H_2} - v|^2 \over  2 \lam^2} 
\\
-  \sqrt{ H_2 \ov{H}_1 \over H_1 \ov{H_2} }
{ |\sqrt{2 H_1 H_2} - v|^2 \over  2 \lam^2} 
&
1 - \left| {H_1 \over H_2} \right|
{ |\sqrt{2 H_1 H_2} - v|^2 \over  2 \lam^2} 
\end{array}
\right) \, 
\label{allowed}
\ee
be positive-definite implies that the allowed field configurations
are those satisfying the inequality
\be
\label{nome}
\sqrt{ {|H_1|^2 + |H_2|^2 \over 2 |H_1||H_2|} } 
\left| \sqrt{2 H_1 H_2} - v \right| < \lam \, .
\ee
Thus the scale $\lam$, which controls the deviations
from a canonical K\"ahler potential, recovered in the
limit $\lam \to \infty$ with $v$ fixed, acts also as
an effective cutoff in field space. The $F$-term
contribution to the scalar potential is
\be
\displaystyle{
||F||^2 = {\lam_S^4 }
{1 \over
\displaystyle{{2|H_1||H_2| \over |H_1|^2 + |H_2|^2}}
-
\displaystyle{{ | \sqrt{2H_1 H_2} - v |^2
 \over \lam^2}} }} \, ,
\label{fcontr}
\ee
and is manifestly positive-definite in the domain of validity of the 
model, so that supersymmetry is spontaneously broken. In terms of the 
usual gauge-invariant variables, the potential has a unique absolute 
minimum, corresponding to $\theta=0$ and  $v_1 = v_2 = v/\sqrt{2}$, 
where $v$ is now the fixed input scale in eq.~(\ref{kfer}).

It is interesting to examine how the mass spectrum and the
interactions are modified with respect to the case of canonical
kinetic terms. The analysis here is very simple, thanks to the
fact that the K\"ahler metric is canonical at the minimum.
In the scalar sector, the Goldstone boson is still $\varphi_A'$,
and the mass of $\varphi_A$ remains the same. However, $\varphi_S$
and $\varphi_S'$ now have non-vanishing masses:
\be
m_S^2 = m_{S'}^2 = {\lam_S^4 \over \lam^2} \, ,
\label{phimm}
\ee
controlled by the supersymmetry-breaking scale and by the curvature
of the K\"ahler manifold at the minimum of the potential. The vector
boson has the same mass as before. In the gaugino--higgsino sector,
the goldstino is still $\tilde{S}$, and the only modification in the 
mass matrix is the possibility of a non-vanishing diagonal gaugino 
entry. For instance, in analogy with eq.~(\ref{fsimple}), we can choose
\be
\label{fsimplebis}
f = {1 \over g^2} \left( 1 + {\sqrt{2 H_1 H_2} -v \over \lam_f} 
\right) \, ,
\ee
where $\Lambda_f$ is an arbitrary mass scale, which generates
a gaugino mass as in eq.~(\ref{gausimple}).

The non-canonical $K$ and $f$ generate new interactions as well,
with coefficients scaling with inverse powers of $\lam$ and $\lam_f$.
In the high-energy r\'egime, these interactions are sources of
unitarity violations, in addition to those generated by the 
superpotential and discussed in section~3.1. Since the unitarity
violations due to $K$ and $f$ are analogous to those discussed
in section~2, we will not repeat the analysis in the present case.
Instead, we will conclude this section with an illustration of
low-energy behaviour. We will reconsider the scattering process 
${\tilde S} \varphi_A \to {\tilde S} \varphi_A$, whose high-energy
behaviour was briefly discussed in section 3.1, using canonical 
kinetic terms and neglecting gauge interactions. We will now discuss 
the same process in the low-energy r\'egime\footnote{We did not
discuss this r\'egime in section~3.1 because the scalars 
$\varphi_S,\varphi'_S$ were massless.}, using the non-canonical 
$K$ and $f$ introduced above, and also including gauge interactions
for completeness. In analogy with the processes ${\tilde S} \lambda \to
{\tilde S} \lambda$ and ${\tilde S} {\tilde A} \to 
{\tilde S} {\tilde A}$ of section~2.3, we will consider the process 
${\tilde S} \varphi_A \to {\tilde S} \varphi_A$ in a Thomson-like 
kinematical limit, in which the incident goldstino has energy 
$E\ll m_A$ in the centre-of-mass frame. Once again, we will see that the
leading individual contributions to the amplitudes, which are proportional 
to $E$, cancel out in the sum, so that the leading energy dependence
is quadratic in $E$. The amplitudes that receive contributions
proportional to $E$ correspond to opposite-helicity goldstinos. More 
specifically, an amplitude of that type receives three classes of 
such contributions, ${\cal A}={\cal A}_1+{\cal A}_2 + {\cal A}_3$, 
which we will  now describe in some detail. 
The first class of contributions arises from the contact interactions 
$\varphi_A^2({\tilde S}{\tilde S} + {\rm h.c.})$ that originate from
eq.~(A.17):
\be
{\cal A}_1 \simeq \left( {2 \lsu^2 \over v^3} + {\lsu^2 \over v \lam^2} 
\right) E \, .
\label{contr1}
\ee
The first term was already present in the canonical case 
(see eq.~(\ref{yukawa}) or eq.~(\ref{contca})), whereas the second 
one is due to the non-canonical $K$. The second class of 
contributions arises from $t$-channel exchange of the $\varphi_S$ 
scalar\footnote{Since $t = {\cal O}(E^2)$, we assume $E \ll m_S$.}: 
\be
{\cal A}_2 \simeq \left( - {4 \lsu^2 \over v^3}-{\lsu^2 \over v \lam^2}  
+ {g^2 v \over \lsu^2 } - {g^2 v^2 \over 2 \lsu^2 \lam_f}
\right) E \, .
\label{contr2}
\ee
As in the examples of section~2, the $m_S^2$ factor in the cubic 
$\varphi_S {\tilde S} {\tilde S}$ coupling (see eq.~(\ref{t3})) is 
cancelled by the $m_S^2$ in the $\varphi_S$ propagator. The above expression 
is proportional to the four terms that contribute to the cubic 
$\varphi_S  \varphi_A^2$ coupling: the first two terms come from $||F||^2$ 
(the second one is due to the non-canonical $K$), and the last two terms 
come from $||D||^2$ (the fourth one is due to the non-canonical $f$).
The third class of contributions arises from $s$- and
$u$- channel exchange of the two mass eigenstates in the fermionic 
${\tilde A}$-$\lambda$ sector\footnote{Since $s \simeq m_A^2 + 2 m_A E$
and $u \simeq m_A^2 - 2 m_A E$, we assume $E \ll |m_1^2-m_A^2|/m_A,
|m_2^2-m_A^2|/m_A$, where $m_1$ and $m_2$ denote the two fermion mass 
eigenvalues.}:
\be
{\cal A}_3 \simeq \left( {2 \lsu^2 \over v^3} 
- {g^2 v \over \lsu^2 } + {g^2 v^2 \over 2 \lsu^2 \lam_f}
\right) E \, .
\label{contr3}
\ee
This expression takes into account both mass mixing effects and 
all the relevant cubic couplings, which are the $\varphi_A {\tilde A} 
{\tilde S}$ and $\varphi_A {\lambda} {\tilde S}$ couplings already 
visible in eqs.~(\ref{yukawa}) and (\ref{gaugeint}), and also an 
additional $\varphi_A {\lambda} {\tilde S}$ coupling that originates
from eq.~(A.19) and is due to the non-canonical $f$. 
Notice that, once the three classes of contributions
(\ref{contr1}), (\ref{contr2}) and (\ref{contr3}) are combined,
an exact cancellation takes place and therefore no linear dependence
on $E$ is present. The leading energy behaviour is only quadratic in $E$: 
it corresponds to scattering amplitudes with equal-helicity goldstinos 
and originates from $s$- and $u$-channel exchange in the fermionic 
${\tilde A}$-$\lambda$ sector. Such amplitudes grow as $E^2 m_A^2/\lsu^4$ 
and lead to cross sections that scale as $\sigma \sim E^4 m_A^2 /\lsu^8$. 
Therefore the low-energy properties of the scattering process ${\tilde S} 
\varphi_A \to {\tilde S} \varphi_A$ are similar to those of the processes
${\tilde S} \lambda \to {\tilde S} \lambda$ and ${\tilde S} {\tilde A} \to 
{\tilde S} {\tilde A}$, discussed in section~2.3. As a final comment,
notice that also the latter processes could be reconsidered in the 
present framework, including gauge interactions and taking into account mass 
mixing effects, but we expect low-energy properties of the same type.

\subsection{Generalizations}

The above $U(1)$ models involving two charged fields and a superpotential 
$w=\lsu^2 \sqrt{2 H_1 H_2}$ can be taken as explicit examples
(i.e. existence proofs) of the simultaneous breaking of supersymmetry 
and gauge symmetry, in the simplest possible setting. We would like
to stress that those examples can be easily extended to models with 
larger gauge groups and, correspondingly, larger Higgs representations. 
The possible physical applications that we have in mind may associate
the broken gauge symmetry with grand $[SU(5),SO(10),E_6,SU(5) \times
U(1),\ldots]$ or electroweak $[SU(2) \times U(1)]$ unification. 
Although in general several analytic gauge invariants can be built, 
we will focus here on straightforward generalizations of the basic 
structure $\sqrt{2 H_1 H_2}$ that emerged in the $U(1)$ case, i.e. we will 
again consider superpotentials that are square roots of gauge invariant 
bilinears in the Higgs fields. We will concentrate on the vacuum and
the basic features of the spectrum, which require only a trivial
generalization of the $U(1)$ Lagrangian presented in the appendix.

\subsubsection{Canonical kinetic terms} 

Consider a generic gauge group $G$ and chiral superfields
in a vector-like representation of $G$, for instance a single 
self-conjugate representation $\Phi$ [vector of $SO(n)$, 
adjoint of $SO(n)$ or $SU(n)$, \ldots] or a conjugate pair 
of complex representations $\Phi_+$ and 
$\Phi_-$. Then, for canonical kinetic terms, superpotentials of the 
(schematic) form $w = \lsu^2 \sqrt{\Phi^2}$ or 
$w = \lsu^2 \sqrt{\Phi_+ \Phi_-}$ 
lead to spontaneous supersymmetry and gauge-symmetry breaking.
To show this, notice that in these cases the Higgs field components can 
be easily rearranged in such a way that 
$w = \lsu^2 \sqrt{\Phi_I \Phi_I}$,
where $I=1,\ldots,N$ runs over the total number of Higgs components.
Thus $w$ has a manifest $SO(N)$ global symmetry, which is also
a symmetry of the canonical $K = \Phi_I {\ov \Phi}_I$ and may 
in general be larger\footnote{In particular, the $\Phi_I$
fields could be just gauge singlets and transform only under the
global $SO(N)$. Then  the focus would shift entirely from
the local to the global symmetry group. This can be
viewed as an intermediate situation between the two scenarios 
discussed in this paper (i.e., goldstino belonging to a neutral or
a charged gauge multiplet) and can be recovered as a special case. 
We recall that global flavour symmetries and singular
superpotentials play an important r\^ole in models of dynamical 
supersymmetry breaking. For instance, ref.~\cite{it} considered a 
superpotential of the above form $w = \lsu^2 \sqrt{\Phi^2}$, with 
$\Phi$ transforming as a singlet under the gauge group and as the 
adjoint of a global $SO(6)$ flavour group, spontaneously broken 
to $SO(5)$.} than the gauge symmetry $G$. This $SO(N)$
symmetry generalizes the $SO(2)\equiv U(1)$ symmetry of the above examples,
where global and local symmetry coincided and $\Phi_I\equiv \{S,iA\}$.
Indeed, the analysis of the vacuum and the spectrum is only slightly
more complicated. To start with, notice again that the $F$-term
contribution to the scalar potential is manifestly positive-definite:
\be
||F||^2 = \lsu^4 \frac{\Phi_I {\ov \Phi}_I}{|\Phi_J \Phi_J|} \geq \lsu^4 \, .
\label{ffss}
\ee
For definiteness, we choose the vacuum where $\Phi_N \equiv S$ is 
the only superfield with non-vanishing scalar (and therefore also 
$F$-component) 
vev\footnote{The choice will be such that $\langle ||D||^2 \rangle =0$ 
as well.}: $\langle S \rangle = v$, $\langle F^S \rangle = \lsu$. 
Therefore the fermion ${\tilde S}$ is the goldstino. Since the
potential is $S$-flat, $v$ is undetermined and the associated scalars
${\varphi}_S$ and ${\varphi'}_S$ remain massless. 

The rest of the spectrum can be conveniently described in terms of the 
(simultaneous) supersymmetric Higgs and Goldstone mechanisms, which are 
in turn corrected by the effects of spontaneous supersymmetry breaking.
We start with the (broken) supersymmetric Goldstone mechanism, 
associated with the global $SO(N) \to SO(N-1)$ breaking and affecting 
all the $N-1$ supermultiplets $\Phi_I$ different from $S$. 
In each multiplet, one real scalar is a massless Goldstone boson,
the other real scalar gets a mass squared contribution equal to 
$2\lsu^4/v^2$, whereas the fermion gets a mass equal to $\lsu^2/v$. 
However, a number $n_B \leq N-1$ of these multiplets are affected at 
the same time by the Higgs mechanism, associated with the breaking of 
the gauge group $G$ to a subgroup $H$ ($n_B={\rm dim}G-{\rm dim}H$).
Let us denote by $V^a=(A^a_{\mu},\lambda^a)$ and 
$\Phi_a=({\varphi}_a,{\varphi}'_a,{\tilde \Phi}_a)$ 
($a=1,\ldots,n_B$) the vector and chiral multiplets associated with 
the $n_B$ broken generators.  
The real scalars ${\varphi}_a$ are unphysical Goldstone bosons and 
are absorbed by the gauge bosons $A^a_{\mu}$, which get 
masses $m_{V_a}\sim g_a v$. Therefore only $N-1-n_B$ of the
original $N-1$ Goldstone bosons remain in the physical 
spectrum. Furthermore, the real scalars ${\varphi}'_a$ 
get $m^2_{V_a}$ mass terms from $||D||^2$, in addition to the 
above mentioned $2\lsu^4/v^2$ term from $||F||^2$. 
Finally, the corresponding fermions ${\tilde \Phi}_a$ get coupled 
to the gauginos $\lambda^a$ via off-diagonal $m_{V_a}$ mass terms, 
besides having the above mentioned $\lsu^2/v$ diagonal mass. 

\subsubsection{Embedding in supergravity}

Similarly to the cases discussed in previous sections,
also the more general models under consideration admit a
simple and interesting embedding in supergravity models
with spontaneously broken supersymmetry and vanishing vacuum 
energy. Adapting the analysis of \cite{bfz} to the present case, 
we find that an underlying supergravity model can be defined
via the $SO(2,N)/SO(2)\times SO(N)$ K\"ahler potential
\be
K = - {3 \over 2} \mpl^2 \log \left[ 1 -
{2 \Phi_I {\ov \Phi}_I \over 3 \mpl^2}  +
{(\Phi_I \Phi_I) ({\ov \Phi}_J {\ov \Phi}_J) \over 9 \mpl^4} \right] \, 
\ee
and the superpotential
\be
w = { \mpl \lam_S^2 \over \sqrt{3}}
\left( 1 +  \sqrt{ {\Phi_I \Phi_I \over 3 \mpl^2} } \right)^3 \, ,
\ee
which generalize (\ref{ksugra}) and (\ref{wsugra}), respectively.
Notice that $w$ and $K$ have a manifest $SO(N)$ symmetry, similarly to the 
$w$ and $K$ of the globally supersymmetric model, which can be recovered 
in the flat limit. The main result is a generalization of
(\ref{vfull}) and (\ref{vappr}), i.e. we find
\bea
||F||^2 - ||H||^2
& = & \Lambda_S^4 
\frac { \left( 1 + { |\Phi_I \Phi_I| \over 3 \mpl^2} \right) \,
\left| 1 +  \sqrt{\Phi_I \Phi_I \over 3 \mpl^2} \right|^4 }
{ \left[ 1 - {2 \Phi_I {\ov \Phi}_I \over 3 \mpl^2}  +
{(\Phi_I \Phi_I) ({\ov \Phi}_J {\ov \Phi}_J) \over 9 \mpl^4} \right]^{3/2} }
\left( \frac{\Phi_I {\ov \Phi}_I }{|\Phi_J \Phi_J|} -1 \right) 
\\
& = & \Lambda_S^4 
\left( \frac{\Phi_I {\ov \Phi}_I }{|\Phi_J \Phi_J|} -1 \right) 
\left[ 1 + {\cal O} \left( {\Phi^2 \over \mpl^2} \right)
\right] \, .
\eea
Therefore both the full $||F||^2-||H||^2$ and the corresponding 
leading term in the $\mpl \to \infty$ expansion are positive
semi-definite and vanish identically along the same minima of
(\ref{ffss}).

\subsubsection{Non-canonical kinetic terms}

We have seen that the globally supersymmetric models with 
canonical kinetic terms described above have two types
of massless scalars. The first type consists of the
scalars $({\varphi}_S, {\varphi'}_S)$, partners of the goldstino
and associated with the (non-compact) complex flat direction
$S$ of the scalar potential. The second type consists
of Goldstone bosons associated with (compact) flat non-gauge 
directions, which are present when the gauge symmetry has
a smaller number of broken generators than the global one.
Mass terms for such scalars, if required for some reasons,
can be generated by appropriate modifications of the model. 
For example, the flat direction $S$ can be lifted with
the following modification of the canonical K\"ahler potential,
which generalizes (\ref{kfers}) and (\ref{kfer}): 
\be
K =  \Phi_I {\ov \Phi}_I  -
{ |\sqrt{ \Phi_I \Phi_I} - v|^4 \over  4 \lam^2} \, ,
\label{kferg}
\ee
where $v$ and $\lam$ are real and positive mass parameters.
The condition that the corresponding K\"ahler metric be
positive-definite implies $\det K_{{\ov M} N}  > 0$. 
Therefore we obtain the following necessary condition
on the allowed field configurations, which generalizes
(\ref{nome}):
\be
\sqrt{ { \Phi_I {\ov \Phi}_I \over 
|\Phi_J \Phi_J|} } \left| \sqrt{ \Phi_I \Phi_I} - v \right| 
< \lam \, .
\ee
The $F$-term contribution to the scalar potential is positive-definite 
in the allowed domain and therefore breaks supersymmetry:
\be
\displaystyle{
||F||^2 = \lsu^4  
{1 \over
\displaystyle{  { |\Phi_I \Phi_I| \over \Phi_J {\ov \Phi}_J } }
-
\displaystyle{{ | \sqrt{ \Phi_I \Phi_I } - v |^2
 \over \lam^2}} }} \geq \lsu^4 \, ,
\ee
much as in (\ref{fcontr}).
The flat direction $S$ is now lifted, since $\langle S \rangle = v$
is determined in terms of the fixed input scale $v$ of eq.~(\ref{kferg}),
and $\varphi_S$ and $\varphi_S'$ now have non-vanishing masses, 
\be
m_S^2 = m_{S'}^2 = {\lam_S^4 \over \lam^2} \, ,
\ee
as in (\ref{phim}) and (\ref{phimm}).

\subsubsection{Standard Model application}

We conclude with an interesting example within the general class
of models described above, i.e. we consider the Standard Model 
gauge group $SU(3)\times SU(2) \times U(1)$ and the usual Higgs 
doublets $H_1 \sim (1,2,-1/2)$ and $H_2 \sim (1,2,+1/2)$.
Then, for canonical kinetic terms, the choice
$w=\Lambda_S^2 \sqrt{2 H_1 H_2}=\Lambda_S^2 
\sqrt{2 (H_1^0 H_2^0- H_1^- H_2^+)}$ leads to the spontaneous breaking 
of supersymmetry, of the gauge symmetry to $SU(3)\times U(1)_{e.m.}$
and of the global symmetry $SO(4)\simeq SU(2)_L \times SU(2)_R$
to the custodial $SO(3)\simeq SU(2)_V$. 
After a convenient choice of phases for the vacuum, the symmetric 
combination of $H_1^0$ and $H_2^0$ plays again the r\^ole of goldstino 
superfield $S$, whereas the antisymmetric combination of $H_1^0$ and 
$H_2^0$ and the charged components $H_1^-$ and $H_2^+$ play the 
r\^ole of $A$-type superfields. Again, since the scalar potential
is flat in the $S$ direction, $\langle S \rangle=v$ is undetermined and
the two neutral Higgs bosons that are partners of the goldstino remain 
massless: $m_S=m_{S'}=0$. The unphysical neutral and charged Goldstone 
bosons are absorbed via the Higgs mechanism by the $Z$ and $W$ gauge bosons,
which get masses $m_Z^2=(g^2+g'^2)v^2/2$ and $m_W^2=g^2 v^2/2$.
The remaining neutral and charged Higgs bosons have masses $m_A^2= 
m_Z^2+2\lsu^4/v^2$ and $m_{H^\pm}^2 = m_W^2+2\lsu^4/v^2$. Since the 
gauge and global symmetry breaking involve the same number of
broken generators (three), no physical massless Goldstone
bosons appear in the spectrum. Moreover, the masslessness of the 
Higgs bosons $(\varphi_S, \varphi_S')$ can be easily cured by 
modifying the canonical K\"ahler potential in such a way 
that $\langle S \rangle$ is no longer undetermined.
For instance, we can choose a modified K\"ahler potential of the 
form (\ref{kfer}), so that the flatness is lifted and  
$m_S^2=m_{S'}^2=\lsu^4/\lam^2$. In particular, both the example with
canonical $K$ and this generalization show that some celebrated
mass sum rules of the MSSM are violated already at the classical
level\footnote{We recall that, in our notation, the two neutral 
$CP$-even Higgs bosons are $\varphi_S$ and $\varphi_A$, while the 
$CP$-odd one is $\varphi'_S$.}: $m_S^2 + m_A^2 \ne m_Z^2 + m_{S'}^2$, 
$m_{H^\pm}^2 \ne m_W^2 + m_{S'}^2$. 
As for the spin-1/2 spectrum, we just recall that one of the four
neutralinos is necessarily massless (the goldstino ${\tilde S}$) and 
that gaugino masses can be generated using a non-canonical gauge 
kinetic function.
Finally, the model can be completed with the addition of quark and 
lepton superfields, which are also $A$-type fields. 
In particular, an appropriate (non-canonical) choice for the 
corresponding K\"ahler metric can easily lead to realistic squark 
and slepton masses.

It is not yet clear if an `ultra-minimal' supersymmetric standard model 
of this type can be made fully realistic. We postpone a detailed
analysis to future work and recall here only the main issues.
On the one hand, the model should satisfy the requirements of tree-level 
unitarity, at least at the present collider energies, and be compatible with 
the absence of measurable deviations in many processes involving
Standard Model particles: schematically, $\Delta m^2 \simlt \lam_S^2$ 
and $\lam_S^2 \simlt G_F^{-1}$. On the other hand, the model 
should be compatible with the non-observation of supersymmetric 
particles and Higgs bosons: schematically, $\Delta m^2 \simgt
G_F^{-1}$. Satisfying such simultaneous requirements represents the
main challenge. However, if the new, non-renormalizable interactions 
are mostly associated with the supersymmetric Higgs sector, which
contains the goldstino, then the model may still survive the known 
phenomenological constraints in some region of its parameter space,
whilst predicting important deviations from the MSSM in the Higgs 
and sparticle properties, some of which could be tested soon.
\vfill{
\section*{Acknowledgements}
We would like to thank S.~Ferrara, C.~Kounnas, A.~Masiero, M.~Porrati, 
R.~Rattazzi, A.~Riotto, A.~Rossi for useful discussions, 
and S.~Rigolin for help with the figures.
}
\newpage
\appendixA{Appendix}
We recall here the explicit form of a Lagrangian with $N=1$,
$d=4$ global supersymmetry and gauge group $G = U(1)$, 
specializing the general formulae of \cite{wb}. We
disregard the possibility of a Fayet--Iliopoulos term.
We use, unless otherwise stated, the conventions of \cite{wb},
apart from the choice of the flat space-time metric and of the
Pauli matrices, taken here to be $\eta_{\mu \nu} = {\rm diag} 
\, (1,-1,-1,-1)$, $\sigma^{\mu} = (1,\vec{\sigma})$,
$\ov{\sigma}^{\mu} = (1,- \vec{\sigma})$. We denote
by $(\varphi^i,\psi^i)$ the physical spin-$(0,1/2)$ components of the 
chiral superfields (denoted here with the same symbols as 
their scalar components) and by $(\lambda,A_{\mu})$ 
the physical spin-$(1/2,1)$ components of the $U(1)$ vector 
superfield $V$. We denote by $Y^{(i)}$ the charges of the different 
chiral supermultiplets. Neglecting as usual higher-derivative terms, 
the defining functions of a model are the following: the K\"ahler
potential $K(\varphi,\ov{\varphi})$, a real gauge-invariant
function of the chiral superfields and their conjugates, of
dimensionality $($mass$)^2$; the superpotential $w(\varphi)$,
an analytic gauge-invariant function of the chiral superfields,
of dimensionality $($mass$)^3$; the gauge kinetic function $f
(\varphi)$, a (dimensionless) analytic gauge-invariant function
of the chiral superfields. When taking derivatives of a generic
function $P(\varphi,\ov{\varphi})$ with respect to the fields,
we adopt the standard conventions
\be
P_i \equiv {\partial P \over \partial \varphi^i} \, ,
\;\;\;\;\;
P_{\ov{\i}} \equiv {\partial P \over \partial
\ov{\varphi}^{\ov{\i}} } \, .
\ee
Moreover, we raise and lower chiral field indices by the action of 
the K\"ahler metric and its inverse, considered as matrices,
with the indices ordered as in $K_{\ov{\j} i}$ and in
$(K^{-1})^{i \ov{\j}}$, so that
\be
\displaystyle{
K_{\ov{\i} k} (K^{-1})^{k \ov{\j}} = \delta_{\ov{\i}}^{\ov{\j}} \, ,
\;\;\;\;\;
(K^{-1})^{i \ov{k}} K_{\ov{k} j} = \delta^i_j \, . }
\ee
For instance, the indices of the auxiliary components 
$F^i \equiv - \varphi^i |_{\theta \theta}$ 
of the chiral superfields can be lowered as follows:
\be
F_i =  K_{\ov{\j} i} \, {\ov F}^{\ov{\j}} \, .
\ee

In superfield notation, the general form of the Lagrangian is
\be
{\cal L} = {1 \over 4}
\left[
\int d^2 \theta f(\varphi) {\cal W} {\cal W} + {\; \rm h.c.} \right]
+
\int d^4 \theta K \left( e^{2V} \varphi,\ov{\varphi} \right)
+ \left[ \int d^2 \theta \, w \left( \varphi \right) + {\; \rm h.c.}
\right] \, ,
\ee
where ${\cal W}$ is the supersymmetric gauge field strength.
As usual, the component Lagrangian for the physical degrees 
of freedom can be obtained by first choosing the Wess--Zumino 
gauge and then eliminating the auxiliary fields through the 
corresponding equations of motion, which consist of a bosonic 
and a fermionic part. We will write explicitly the fermionic 
terms in the Lagrangian and reserve the symbols $F_i$, $D_Y$ 
for the bosonic parts of the auxiliary fields:
\bea 
F_i & = &  w_i  \, ,
\\
D_Y & = & Y^{(i)} K_i \varphi^i  = Y^{(i)}
{\ov \varphi}^{\ov{\i}} K_{\ov{\i}} \, .
\eea
The Lagrangian for the physical fields can be decomposed
as a sum of four parts:
\be
{\cal L} = {\cal L}_B + {\cal L}_{F,K} + {\cal L}_{F,2}
+ {\cal L}_{F,4} \, .
\ee
The purely bosonic part of the Lagrangian is
\bea
{\cal L}_B & = &
- {1 \over 4} ({\rm Re} \, f) F_{\mu \nu}  F^{\mu \nu}
+ {1 \over 8} ({\rm Im} \, f) \epsilon^{\mu \nu \rho \sigma}
  F_{\mu \nu}  F_{\rho \sigma}
\label{bosd}
\\
& &
+ \left( \ov{D^{\mu} \varphi} \right)^{\ov{\j}}
  \, K_{\overline{\j} i} \, \left( D_{\mu} \varphi
   \right)^i - V_{SUSY} \, ,
\label{bosq}
\eea
where $\epsilon_{0123} = - \epsilon^{0123} = 1$ and the scalar 
potential is
\be
\label{sapp}
V_{SUSY} = ||F||^2 + ||D||^2  \, ,
\ee
with 
\be
||F||^2 \equiv
F_i (K^{-1})^{i \ov{\j}} \, \ov{F}_{\ov{\j}}
\, ,
\ee
\be
||D||^2 \equiv
{1 \over 2} ({\rm Re} \, f)^{-1} D_Y^2 \, .
\ee
The part of the Lagrangian that is bilinear in the fermionic fields
and contains space-time derivatives is
\bea
{\cal L}_{F,K} & = &
{i \over 2} ({\rm Re} \, f) \left[
\lambda \sigma^{\mu} ( \partial_{\mu} \ov{ \lambda} )
+
\ov{\lambda} \ov{\sigma}^{\mu} (\partial_{\mu} \lambda)
\right]
- {1 \over 2} ({\rm Im} \, f) \partial_{\mu} \left[
\lambda \sigma^{\mu} \ov{\lambda} \right]
\label{fkd}
\\ & - &
{1 \over 2 \sqrt{2}} \left[ f_{i} \psi^i
\sigma^{\mu \nu} \lambda F_{\mu \nu} + {\rm h.c.} \right]
\label{fkc}
\\ & + &
i \,  K_{\ov{\j} i} \ov{\psi}^{\ov{\j}}
\ov{\sigma}^{\mu} (D_{\mu} \psi)^i
\label{fko}
\, .
\eea
The non-derivative part of the Lagrangian bilinear in
the fermionic fields is
\bea
{\cal L}_{F,2} & = &
{1 \over 4}
f_{i} (K^{-1})^{i \ov{\j}} \ov{w}_{\ov{\j}}
\, \lambda \lambda + {\rm h.c.}
\label{lfu}
\\ & - &
{1 \over 2} \left[ w_{i j} - w_l (K^{-1})^{l \ov{k}}
K_{\ov{k} i j } \right] \psi^i \psi^j + {\rm h.c.}
\label{lfd}
\\ & + &
i \sqrt{2} D_{Y  , i} \psi^i \lambda + {\rm h.c.}
\label{lft} \\ & - &
{i \over 2 \sqrt{2}}
({\rm Re} \, f)^{-1} f_{i}  D_Y \psi^i \lambda + {\rm h.c.}  \, ,
\label{lfc}
\eea
whereas the one containing four-fermion interactions is
\bea
{\cal L}_{F,4} & = &
- {1 \over 8}
({\rm Re} \, f)^{-1} f_{i} \ov{f}_{\ov{\j}} \,
\psi^i \lambda \ov{\psi}^{\ov{\j}} \ov{\lambda}
\label{ffd} \\ & &
+ {1 \over 8}
\left[ f_{ ij} - f_{l} (K^{-1})^{l \ov{k}}
K_{\ov{k} i j } - {1 \over 4} ({\rm Re} \, f)^{-1} f_{i}
f_{j} \right] \psi^i \psi^j \lambda \lambda + {\rm h.c.}
\label{ffq} \\ & &
-{1 \over 16} f_{i} (K^{-1})^{i \ov{\j}} \ov{f}_{\ov{\j}} \,
\lambda \lambda  \ov{\lambda} \ov{\lambda}
\label{ffo} \\ & &
+ {1 \over 4} \left[ K_{i \ov{\j} k \ov{l}}
- K_{\ov{\j} \ov{l} m} (K^{-1})^{m \ov{n}} K_{\ov{n} i k} 
\right] \psi^i \psi^k \ov{\psi}^{\ov{\j}} \ov{\psi}^{\ov{l}} 
\, .
\label{ffn}
\eea
We recall that the action of the gauge-covariant and K\"ahler-covariant
derivative $D_{\mu}$ on the scalar and fermionic components of
the chiral superfields is
\be
(D_{\mu} \varphi)^i = \partial_{\mu} \varphi^i
+ i A_{\mu} Y^{(i)} \varphi^i \, ,
\ee
\be
(D_{\mu} \psi)^i = \partial_{\mu} \psi^i
+ i A_{\mu} Y^{(i)} \psi^i + (K^{-1})^{i \ov{m}}
K_{\ov{m} k l} (D_{\mu} \varphi)^k \psi^l  \, .
\ee

Before concluding, we also recall the explicit form of the scalar
potential in the locally supersymmetric version of the model, in order
to make the connection easier between the two theories in the presence
of low-energy supersymmetry breaking. Again, the defining functions
are the K\"ahler potential $K$, the superpotential $w$ and the gauge
kinetic functions $f$. The scalar potential of $N=1$, $d=4$
supergravity has the general form \cite{sugraym}
\be
\label{app}
V_{SUGRA} = ||F||^2 + ||D||^2 - ||H||^2 \, ,
\ee
where, keeping manifest the dependences on the (reduced) Planck 
mass $\mpl \equiv (8 \pi G_N)^{-1/2} \simeq 2.4 \times 10^{18}
\gev$,
\be
||F||^2 \equiv F_i (K^{-1})^{i \ov{\j}} F_{\ov{\j}} \, ,
\ee
\be
\displaystyle{
F_i = \displaystyle{e^{K \over 2 \mpl^2}} \left( w_i +
{K_i \over \mpl^2} w \right) \, ,
}
\ee
and $||D||^2$ has exactly the same form as in the global case. The
last term in $V_{SUGRA}$ is associated with the auxiliary field $H$
of the gravitational supermultiplet:
\be
||H||^2 \equiv 3 { |w|^2 \over \mpl^2} e^{K/\mpl^2} \, ,
\ee
and its vacuum expectation value is related with the
gravitino mass:
\be
m_{3/2}^2 = { \langle || H ||^2 \rangle \over 3  \mpl^2}
= { \langle |w|^2  e^{K/\mpl^2} \rangle \over \mpl^4} \, .
\ee
Notice that the scalar potential of global supersymmetry, $V_{SUSY}$,
is easily recovered from $V_{SUGRA}$ by removing the gravitational
contribution, $- ||H||^2$, and by taking formally the limit
$\mpl \to \infty$ in the definition of $||F||^2$, so that $ F_i 
\to w_i $. Notice finally that, if the vacuum energy is vanishing:
\be
\label{scales}
\langle ||F||^2 + ||D||^2 \rangle = \langle ||H||^2 \rangle
\Longrightarrow
m_{3/2}^2 = { \langle ||F||^2 + ||D||^2  \rangle \over 3  \mpl^2}
\, .
\ee
Given the present bounds on the cosmological constant (see, for instance,
\cite{cosmo}), $\Lambda_{cosm} \equiv \langle V \rangle^{1/4}
\simlt 10^{-4} \ev$, eq.~(A.31) must be satisfied to an extremely 
good accuracy in any realistic model.
\newpage
\end{document}